\documentclass{aa}
\usepackage{txfonts}
\usepackage{natbib}
\usepackage{graphicx}
\usepackage{amssymb}
\usepackage{lscape}
\usepackage{aalongtable}
\usepackage{mathrsfs}
\usepackage[dvips]{rotating}
\bibpunct{(}{)}{;}{a}{}{,}

\def\Ha{H$\alpha$ }
\def\kms{$\mathrm{km}~\mathrm{s}^{-1}$}

\def\pfd{Pf\,$\delta$~}
\def\Pfd{Pfund\,$\delta$~}
\def\Wm{$\mathrm{W}\, \mathrm{m}^{-2}$}

\begin{document}

\title{A survey for nanodiamond features \break in the 3 micron spectra of Herbig Ae/Be stars\thanks{Based on observations collected
at the European Southern Observatory, La Silla, Chile (program numbers
071.C-0653, 072.C-0648, 073.C-0245 and 074.C-0248).}
} 
\titlerunning{Diamonds in Herbig Ae/Be systems}
\authorrunning{Acke \& van~den~Ancker}
\author{B.~Acke\inst{1,}\thanks{Postdoctoral researcher of the Research
  Fund KULeuven.} \and M.~E.~van~den~Ancker\inst{2}}
\institute{Instituut voor Sterrenkunde, KULeuven, Celestijnenlaan 200B, 
3001 Leuven, Belgium\\ 
\email{bram@ster.kuleuven.be}
\and
European Southern Observatory, Karl-Schwarzschild Strasse 2, D-85748
Garching bei M\"unchen, Germany}
\date{DRAFT, \today}

\abstract{}
{ We have carried out a survey of 60 Herbig Ae/Be stars in the 3
  micron wavelength region in search for the rare spectral features at
  3.43 and 3.53~micron. These features have been attributed to the
  presence of large, hot, hydrogen-terminated nanodiamonds. Only two
  Herbig Ae/Be stars, \object{HD~97048} and \object{Elias~3$-$1} are
  known to display both these features.}
{ We have obtained
  medium-resolution spectra ($R \sim 2500$) with the ESO near-IR
  instrument ISAAC in the 3.15$-$3.65~micron range. } 
{ In our sample,
  no new examples of sources with prominent nanodiamond features in
  their 3~micron spectra were discovered. Less than 4\% of the Herbig
  targets show the prominent emission features at 3.43 and/or
  3.53~$\mu$m. Both features are detected in our spectrum of
  HD~97048. We confirm the detection of the 3.53~$\mu$m
  feature and the non-detection of the 3.43~$\mu$m
  feature in MWC~297. Furthermore, we report tentative 3.53~$\mu$m
  detections in V921~Sco, HD 163296 and T~CrA. The sources which
  display the nanodiamond features are not exceptional in the group of
  Herbig stars with respect to disk properties, stellar
  characteristics, or disk and stellar activity. Moreover, the
  nanodiamond sources are very different from each other in terms of
  these parameters. We do not find evidence for a recent
  supernova in the vicinity of any of the nanodiamond sources. \\ 
  We have analyzed the PAH~3.3~$\mu$m feature and
  the \Pfd hydrogen emission line, two other spectral features which
  occur in the 3~micron wavelength range. We reinforce the conclusion
  of previous authors that flared-disk systems display significantly
  more PAH emission than self-shadowed-disk sources. The \pfd line
  detection rate is higher in self-shadowed-disk sources than in the
  flared-disk systems.}  
{ We discuss the possible origin and paucity of the (nano)diamond
  features in Herbig stars. Different creation mechanisms have been
  proposed in the literature, amongst others in-situ and
  supernova-induced formation. Our data set is inconclusive in proving
  or disproving either formation mechanism.}
\keywords{circumstellar matter --- stars: pre-main-sequence ---
             planetary systems: protoplanetary disks}

\maketitle


\section{Introduction}

Herbig Ae/Be (HAEBE) stars are pre-main-sequence objects of intermediate
(1.5--8 M$_\odot$) mass. These sources exhibit large infrared (IR)
excesses, for the greater part due to thermal emission of circumstellar
dust. The geometry of this dust surrounding the late-B, A and F-type
members of this class is disklike \citep[e.g.][]{mannings97, testi,
pietu, fuente, natta04}. Also for early-B-type stars, evidence for
disks has been found \citep[e.g.][]{vink02}. However, there is growing
evidence that the structure of the circumstellar matter in early-type
B stars is fundamentally different from Herbig Ae and T~Tauri stars
\citep[e.g.][]{fuente,bik04}.

The chemistry and mineralogy of the dust component of circumstellar 
disks around HAEBE stars has been studied with unprecedented precision
based on near-to-mid-IR spectra (2--200~$\mu$m) provided by the
\textit{Infrared Space Observatory} \citep[ISO,][]{kessler}. The
launch of this satellite in 1995 was the start of a brand-new era
for the research of protoplanetary disks. The spectra revealed a large
variety 
in dust properties and species, from small carbonaceous molecules like
polycyclic aromatic hydrocarbon molecules (PAHs) to silicate
dust. Moreover, 
some sources were shown to contain crystalline grains, similar to
those found in comets in our own solar system
\citep{waelkens96,malfait, 
malfait99,vandenancker00a, vandenancker00b, meeus01}. All
available ISO 2--15~$\mu$m spectra of HAEBE stars have been studied as
a whole by \citet[][hereafter AV04]{ackeiso}.

A handful of HAEBE sources \citep[HD~97048 and
Elias~3$-$1,][and references therein]{vankerckhoven} and
the post-AGB binary \object{HR~4049} \citep{geballe89} display
peculiar spectral 
features at 3.43 and 3.53~$\mu$m. Comparison with laboratory spectra
has convincingly shown that the carriers of these features are
hydrogen-terminated nanometer-sized diamonds \citep[hereafter
nanodiamonds,][]{guillois}. Near-IR observations of HD~97048 by
\citet{habart04a} have spatially resolved the emission region of the
strong 3.43 and 3.53~$\mu$m features on a scale of
$0.18^{\prime\prime} \times 0.18^{\prime\prime}$. The observations
prove beyong doubt that the emission region is confined to the inner
15~AU of the circumstellar {\em disk}. More tentative 
detections of the features were reported in the literature
\citep[e.g. in HD~142527 and
HD~100546,][respectively]{waelkens96,malfait98b}, but only one
additional 3.53~$\mu$m feature was discovered persuasively
\citep[in MWC~297,][]{terada}. Summing up, only two HAEBE
stars display both the 3.43 and the 3.53~$\mu$m feature, and one
HAEBE object shows the 3.53~$\mu$m feature. Considering that the ISO
sample contains 45 targets, only a minor fraction of the HAEBE stars
appear to have such a spectacular 3 micron spectrum (AV04).

The currently operating successor to the ISO satellite, Spitzer, is
not equipped with a spectrograph which can observe in the 3~micron
region of the spectrum. Ground-based observations are therefore the
only alternative.
With the present paper, we intend to enlarge the sample of HAEBE stars
that are observed in the 3 micron region and possibly identify new
targets which display nanodiamond emission. Therefore, we have employed
the near-IR ISAAC instrument (ESO$-$VLT) and focused on the wavelength
region around 3.3 (polycyclic aromatic hydrocarbons, hereafter PAHs),
3.43 and 3.53~$\mu$m (nanodiamonds). In 
Sect.~\ref{dataset}, we present the sample and the data reduction
procedure. The detected features are discussed in
Sect.~\ref{observations}.

\section{The data set \label{dataset}}

\subsection{The sample stars}

The sample of HAEBE stars compiled for this survey has been based on
Table~1 and 2 in the paper of \citet{the} and on the list published by
\citet{malfait}. For 19 of the 60 observed targets, including
HD~97048, spectroscopic data 
in the 3~micron region are present in the ISO data archive (AV04). The 
overlap between the two samples guarantees the continuity between the
two data sets. This survey enlarges the number of high-quality
3~micron spectra of Herbig stars available to us from 45 (ISO) to 82
(ISO$+$new observations).

\subsection{Observations and data reduction}

The observations in the 3 micron region of the 
spectrum were performed using the medium-resolution spectroscopic mode
of the ESO$-$VLT 
instrument ISAAC\footnote{http://www.eso.org/instruments/isaac/},
mounted on UT1 (ANTU). For each target two wavelength regions were 
recorded: the first is centered around 3.3 and the second around
3.5~$\mu$m. The wavelength coverage of each spectrum is about
0.25~$\mu$m. The width of the unresolved telluric lines in the spectra
is a measure for the resolution of the spectra. This value is
approximately 15\r{A}, corresponding to a spectral resolution of
2300. The spectra are oversampled with a pixel-to-pixel
wavelength step of 2.5\r{A}. 
The width of the slit is 1\arcsec, needed to 
achieve the medium spectral resolution. The sky background subtraction
is done by chopping/nodding with a throw of 20\arcsec~in
the direction of the slit. 

The 3 micron region is heavily affected by atmospheric absorption
features. Standard stars (STDs) have been measured in order to correct
for this effect. The straightforward data reduction procedure consists of a
flatfield correction, wavelength calibration and multiplication with a
responsivity curve. The latter is deduced from the STD
spectrum, scaled to the right flux values using an appropriate Kurucz
model. It was noted that the wavelength calibration of the target and
standard star is never exactly the same. The offset, a fraction of
a pixel up to a few pixels large, leads to
artefacts in the resulting spectrum. We have applied a relative shift
to the extracted STD spectrum to correct for this effect,
in an attempt to reduce the residuals of atmospheric features.

During a few nights, no STD measurements were available to
correct for the telluric line absorption. Some more nights, 
flatfield and/or wavelength calibration files were lacking\footnote{No
STD in the 3.3~$\mu$m region on 04--05/09/2003 and 06--07/10/2004; no
STD in both the 3.3 and 3.5~$\mu$m regions on 18--19/11/2003 and
17--18/07/2004; no calibration files on 04--05/11/2003 and
20--21/10/2004}. In these cases, STD and calibration
measurements from other nights were used. This increases the problem
of wavelength calibration significantly, introducing second-order
differences between the target and the STD spectrum.
Furthermore, the assumption that the atmospheric conditions are
approximately the same when observing target and standard star through
the same airmass is not valid at all in these cases. We have tried to 
correct for 
the first order effect of this problem by applying an additional
factor ($\sim$0.8--1.2) to the airmass ratio of target and standard
star. This cosmetic surgery can of course not compensate for
night-to-night differences in e.g. humidity. 

The 3.3 and 3.5~micron spectra were reduced independently. When
comparing the flux levels in the overlap region, 
flux-level jumps occur for a few sources. This is probably due to slit
losses: the slit 
width is only a fraction of the atmospheric seeing. Variations in
seeing between the observations of standard star and target make that
a different fraction of the source's flux is captured. This effect
translates to a multiplicative factor in the resulting spectrum.
Other possible causes for the flux-level jumps include errors in
background subtraction or differences in 
atmospheric conditions between the observations at 3.3 and 3.5~micron.
We have corrected the jump between the two resulting spectra by
applying a multiplicative factor to both spectra and scale them to the
mean value in the overlap region. In most cases, this factor is
of the order of 0.6--1.5. We note that, since the
PAH~3.3~micron and nanodiamond 3.53~micron feature 
are both included within one spectrum, their shape is
insensitive to offsets in absolute flux level. The strength of these
features is of course altered due to the application of the
multiplicative factor.

\clearpage
\begin{figure*}
\rotatebox{0}{\resizebox{\textwidth}{!}{\includegraphics{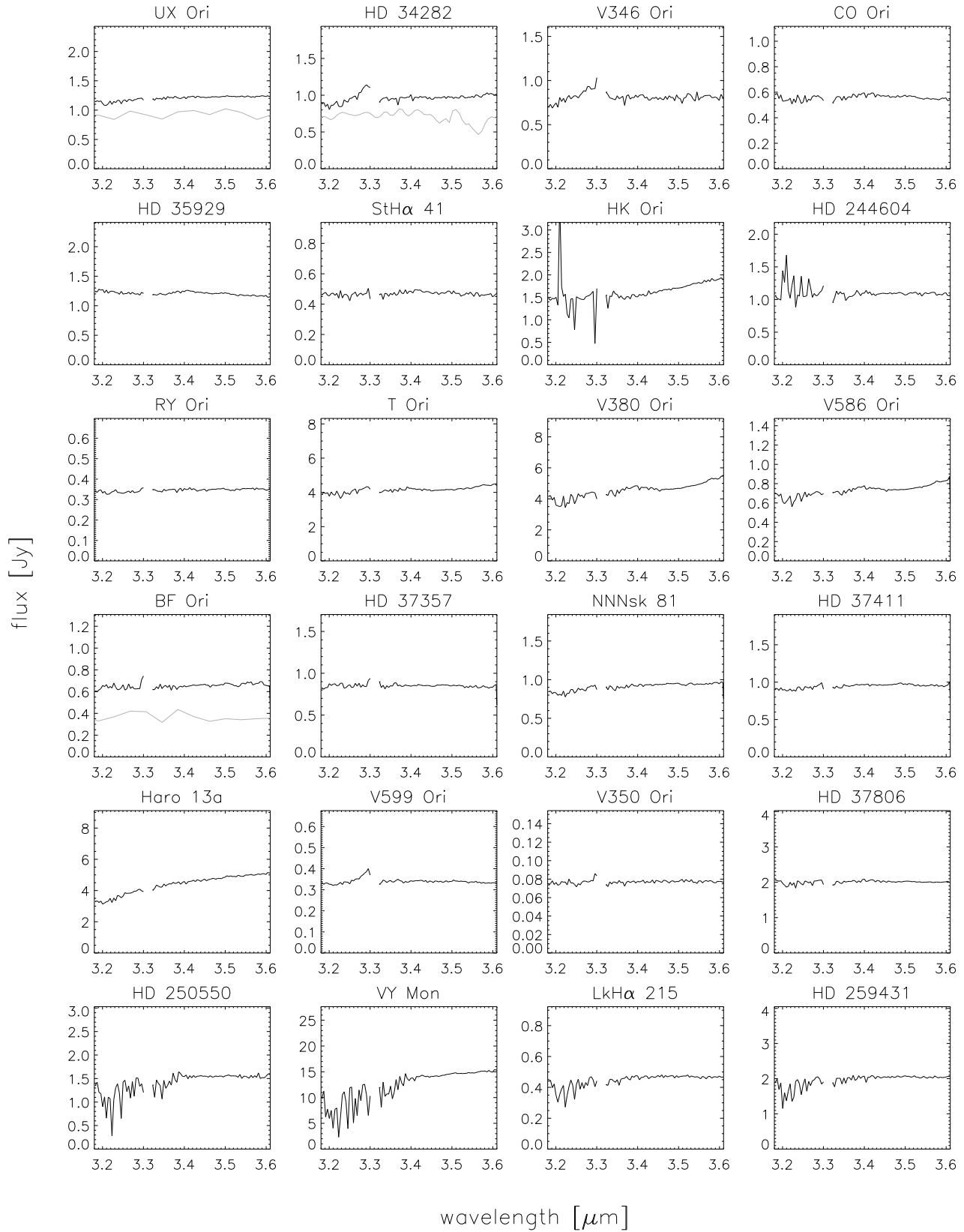}}} 
\caption{ The ISAAC spectra of the sample stars, rebinned to a
  resolution of 750. When available, ISO$-$SWS ($R = 150$) or
  PHT$-$S ($R \sim 90$) spectra are overplotted in
  grey. The part of the ISAAC spectrum between 3.3 and
  3.32~$\mu$m has been excluded. This part is
  unreliable due to the high atmospheric opacity in this wavelength
  range.}
\label{specR750_1.ps}
\end{figure*}

\clearpage
\addtocounter{figure}{-1}
\begin{figure*}[!h]
\rotatebox{0}{\resizebox{\textwidth}{!}{\includegraphics{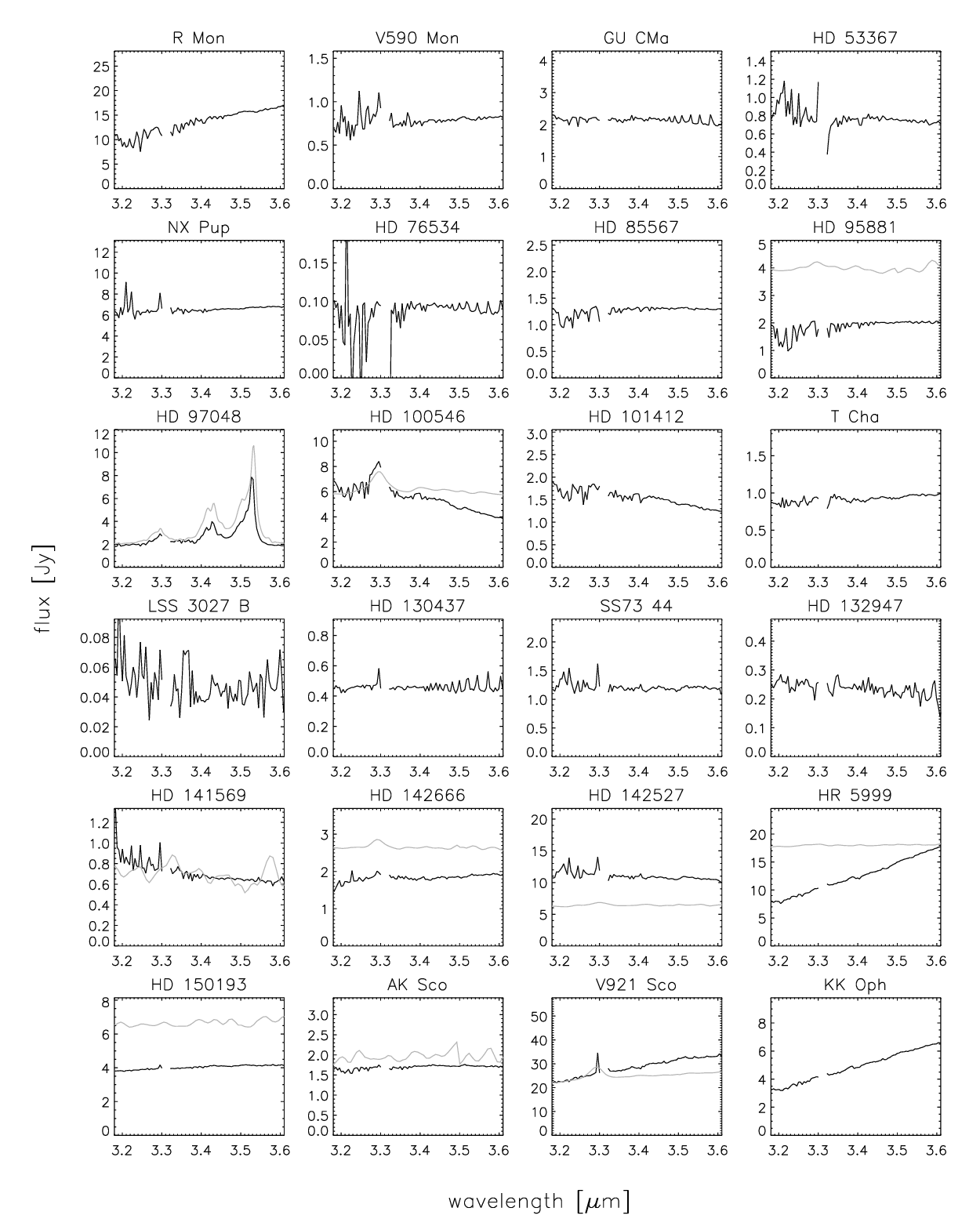}}} 
\caption{ --- (Continued)}
\end{figure*}

\clearpage
\addtocounter{figure}{-1}
\begin{figure*}[!h]
\rotatebox{0}{\resizebox{\textwidth}{!}{\includegraphics{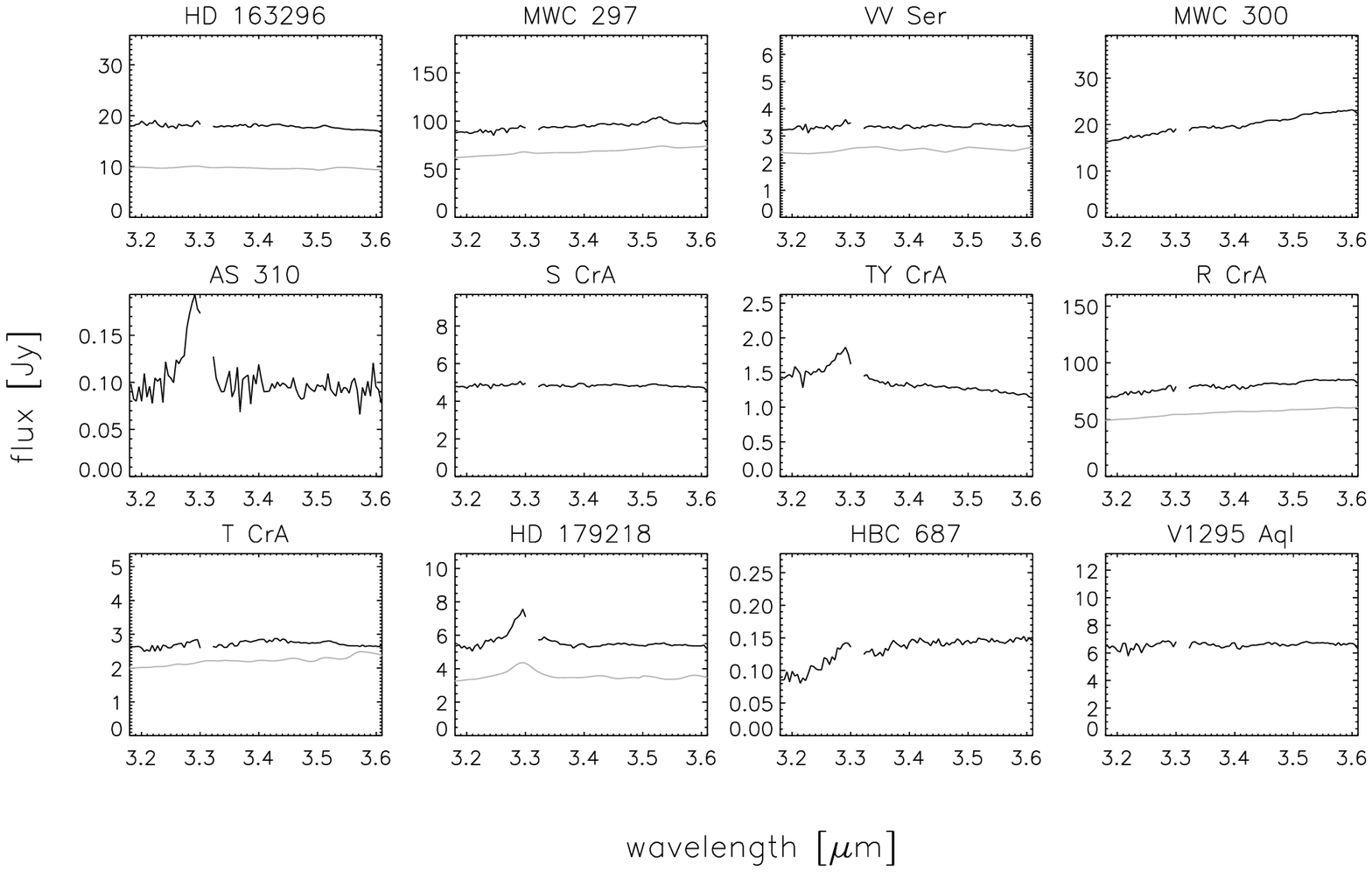}}} 
\caption{ --- (Continued)}
\end{figure*}
\clearpage

Inherent to the reduction method, some residuals of telluric lines
remain present in the final spectra. This is nevertheless not dramatic
for our survey. Because of the medium spectral resolution of our ISAAC
spectra ($\sim$2300), the telluric-line residuals are very narrow
compared to the typical width (FWHM$\sim$0.03$\mu$m) of the PAH and
nanodiamond features. Furthermore a bright 3.53~$\mu$m source similar
to HD~97048 would be easily recognized, even without atmospheric 
corrections, due to the strength of the nanodiamond features.

The reduced ISAAC~3~micron spectra of the sample sources are presented
in Fig.~\ref{specR750_1.ps}. We have rebinned the spectra
to a spectral resolution of 750 for the sake of clarity. The broad PAH
and diamond features remain clearly visible. The sharp, unresolved
\Pfd~line on the other hand is spread out over 20 pixels and is
therefore undetectable in most of the spectra shown in the figure. The
smooth+rebin procedure makes that also the unresolved 
telluric-line residuals are spread out. In regions which are rich in
telluric absorption lines, this may cause problems: the lines are
overlapping each other and form absorption {\em bands} in the rebinned
spectrum. This induces ``absorption features''. It is clear that, of
the two wavelength regions, the 3.3~$\mu$m spectra suffer the most
from telluric line absorption. The region between 3.3 and 3.32~$\mu$m
has been cut out. The measured flux in this
region is close to zero in both science and standard-star
observations, due to strong telluric absorption. Division by the STD
spectrum induces a strong artefact in this wavelength interval. Note
however that the PAH 3.3~$\mu$m feature can still be discerned, since
the onset and peak position of its spectral signature are located at 
shorter wavelengths.

When available, the ISO spectrum is overplotted in
Fig.~\ref{specR750_1.ps}. The continuum flux levels of the ISAAC and
ISO data are within a 30\% error bar of the mean value of both for
most targets. The
experimental error bar on the ratio of the ISAAC continuum flux and
the L-band magnitude ---converted to the right units--- is about
60\%. The latter indicates the poor absolute flux calibration quality
of the ISAAC spectra.

\subsection{Characterizing the spectral features}

Spectral features that appear in the 3 micron spectra are in this
analysis characterized with a few parameters. The integrated line flux
(in \Wm), full width at half maximum (FWHM, in $\mu$m), the peak flux 
(in Jy) and peak position (in $\mu$m) are 
determined. We have estimated the statistical error on these
measurements based on the noise on the spectra. The error on the line
fluxes due to uncertainties on the absolute flux level calibration is
estimated to be 60\%. This error is not included in the figures and
tables presented in this paper. Note that the shape and the
peak-over-continuum ratio of the detected features is not affected
by the absolute calibration of the spectra. In
Sect.~\ref{observations} we discuss the observed spectral features and
include a comparison with the ISO measurements described in AV04.

\subsection{Classifying the spectral energy distributions}

\begin{figure}
\rotatebox{0}{\resizebox{3.5in}{!}{\includegraphics{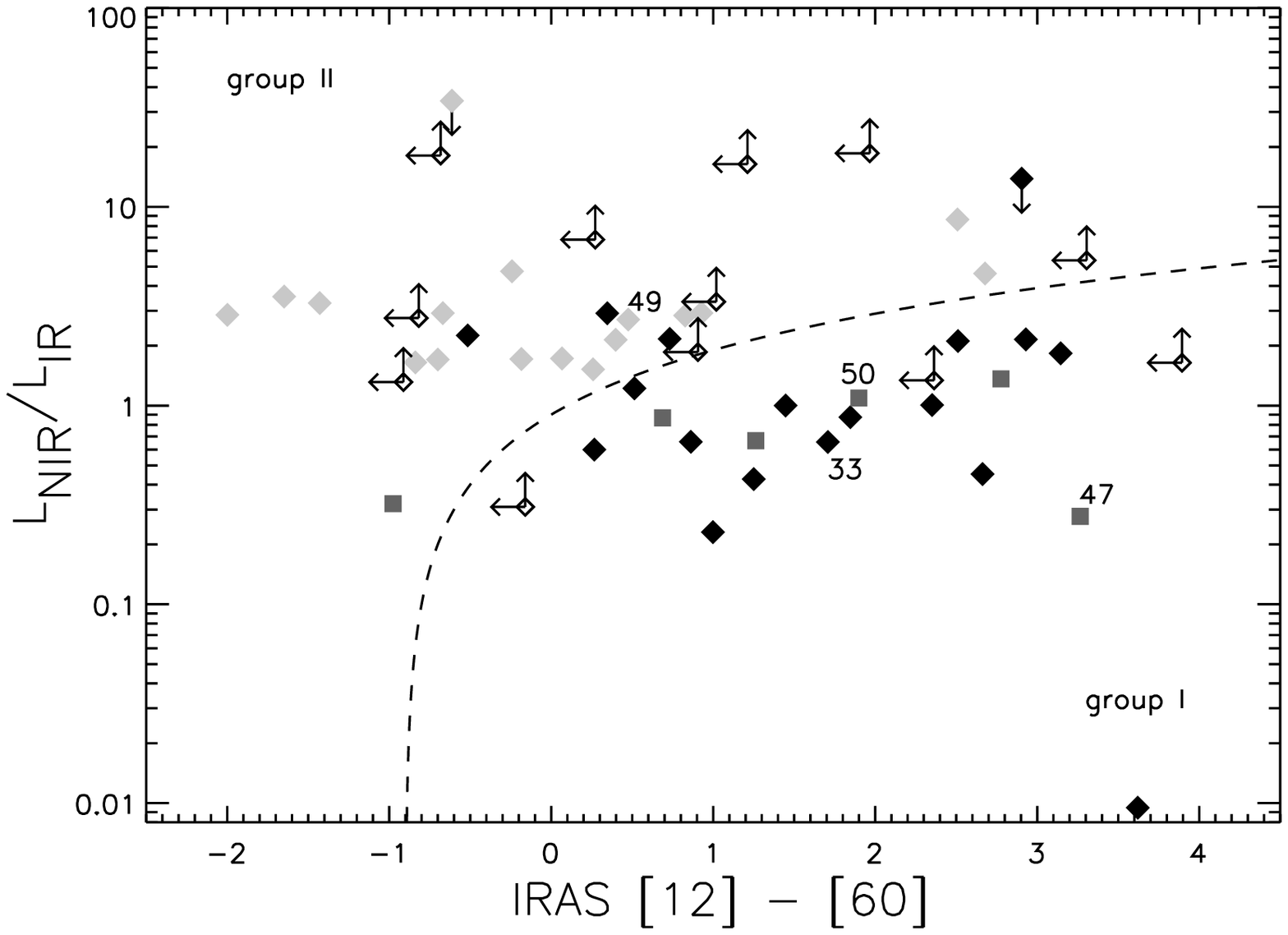}}} 
\caption{ The classification diagram of the sample targets. Sources
  with $A_V$ values above 4 are definied to be embedded sources (grey
  squares). Objects on the left side of the empirical separation line
  (dashed line) are group~II members, on the right side group~I. A
  classification solely based on the 60 micron excess gives almost
  equal results. Objects with a 60 micron excess exceeding 10~mag
  (black diamonds) are almost all group~I sources, while targets with
  an excess below 10~mag (grey diamonds) belong mostly to
  group~II. Some targets have been marked with their sequence number
  (see Table~\ref{presoffeat}); those are the (tentative) nanodiamond
  sources (see Sect.~\ref{sectnano}).}
\label{classification.ps}
\end{figure}

We have classified the sample targets into the two \citet{meeus01}
groups based on the shape of the spectral energy distribution
(SED). This classification roughly represents the circumstellar 
disk's geometry: group~I sources have a strong mid-IR excess which is
believed to originate from a flared outer disk. Group~II objects have a
smaller mid-IR excess and non-flaring outer disks
\citep{dullemond02}.
The sample targets in this paper are classified
based on Fig.~\ref{classification.ps} \citep[see
also][AV04]{vanboekel, dullemond03,ackesubmm}. In this diagram, the
ratio of near-IR and mid-IR luminosity is plotted versus the
non-color-corrected IRAS [12]-[60] color. Group~I sources and embedded
objects end up in the lower right corner of the diagram, while
group~II sources have higher near-IR to mid-IR luminosity ratios. 

Sources with an $A_V$ value exceeding 4 have been classified as
embedded sources. The link between SED shape and circumstellar
geometry is less clear in these cases. In
Fig.~\ref{classification.ps}, we have also included another possible
classification criterium. Almost all group~I and all embedded sources
have a 60 micron excess exceeding 10~mag. In group~II most objects
have an excess smaller than 10~mag. Both criteria ---the diagram and
the 60 micron excess--- hence lead to more or less the same
classification. In Table~\ref{presoffeat} we have indicated the
classification based on the diagram.

Two sample stars, T~CrA and T~Ori, cannot be classified due to a lack
of IR photometric measurements. However, both sources display
photometric variability: their optical light curves are characterized
by Algol-like minima, commonly referred to as UX Orionis (UXOR) 
behaviour. These minima are attributed to extinction events which are
caused by dust and gas clumps moving in and out the line of sight
\citep{wenzel68,grinin88,voshchinnikov90,grinin98}. \citet{dullemond03}
have argued that the clumps spring up from the puffed-up inner rim, and
that all UXORs have self-shadowed disks. In other words, all UXOR
stars should be group II members. The authors indeed find
this correlation in a large sample of Herbig stars. Based on this
conclusion, we classify the UXOR sample targets T~CrA and T~Ori as
group~II members.

In the present sample, 15 sources have been classified as group~I
members, 32 as group~II sources and 6 as embedded objects. 7 sources
remain unclassified due to a lack of photometric measurements and/or
upper limits.


\section{Description of the detected features \label{observations}}

In Table~\ref{presoffeat} we summarize the detection of features in
the 3 micron spectra of the sample stars. A detection is defined as
being a measurement of a spectral feature with a line flux exceeding
3$\sigma$. Tentative detections are features with line
fluxes between 1 and 3$\sigma$. We have searched for the
narrow Pfund~$\delta$ line at 3.296~$\mu$m, the PAH 3.3~$\mu$m feature
and the nanodiamond features at 3.43 and
3.53~$\mu$m. In five sources, also the Humphreys series of the
hydrogen atom was detected. 

\subsection{The Pfund \protect$\delta$ emission line}

The \Pfd line at 3.296~$\mu$m is located at the longer wavelength end
of a region which is rich in telluric lines. The line is
spectrally unresolved in all cases (i.e. FWHM of the order 
of 15\r{A} or equivalently 130~\kms). We have determined the
characterizing parameters of the \pfd line at the full resolution of
the ISAAC spectra (R$=$2300). The error bars on
the line flux are consequently rather large. They are calculated from
the noise in the 3.3 micron region of the reduced spectrum. The error
includes the photon noise, as well as the noise induced by telluric
line residuals. The latter is by far more important.

In the ISAAC
sample of 60 sources, we detected the \pfd emission line 
in 16 targets or 27\%. Four more objects tentatively display this
feature, which adds up to a detection rate of 33\% in this
sample. 

The \Pfd line has been (tentatively) detected in only three
group~I sources (HD~259431, HD~100546 and HD~142527). The detection
rate in this group hence is 13$-$20\%. In 
group~II, 10$-$13 sources (tentatively) display the feature
(31$-$41\%), while half of the six embedded sample stars show the
emission line (50\%). In the group of unclassified sample stars, the
detection rate is 14\%.
The \Pfd line is more detected in group~II and in the
embedded sources, than in group~I. This difference is not
statistically significant on a 5\% level.

\subsection{The Humphreys series \label{secthum}}

\begin{figure*}
\resizebox{\textwidth}{!}{\rotatebox{0}{\includegraphics{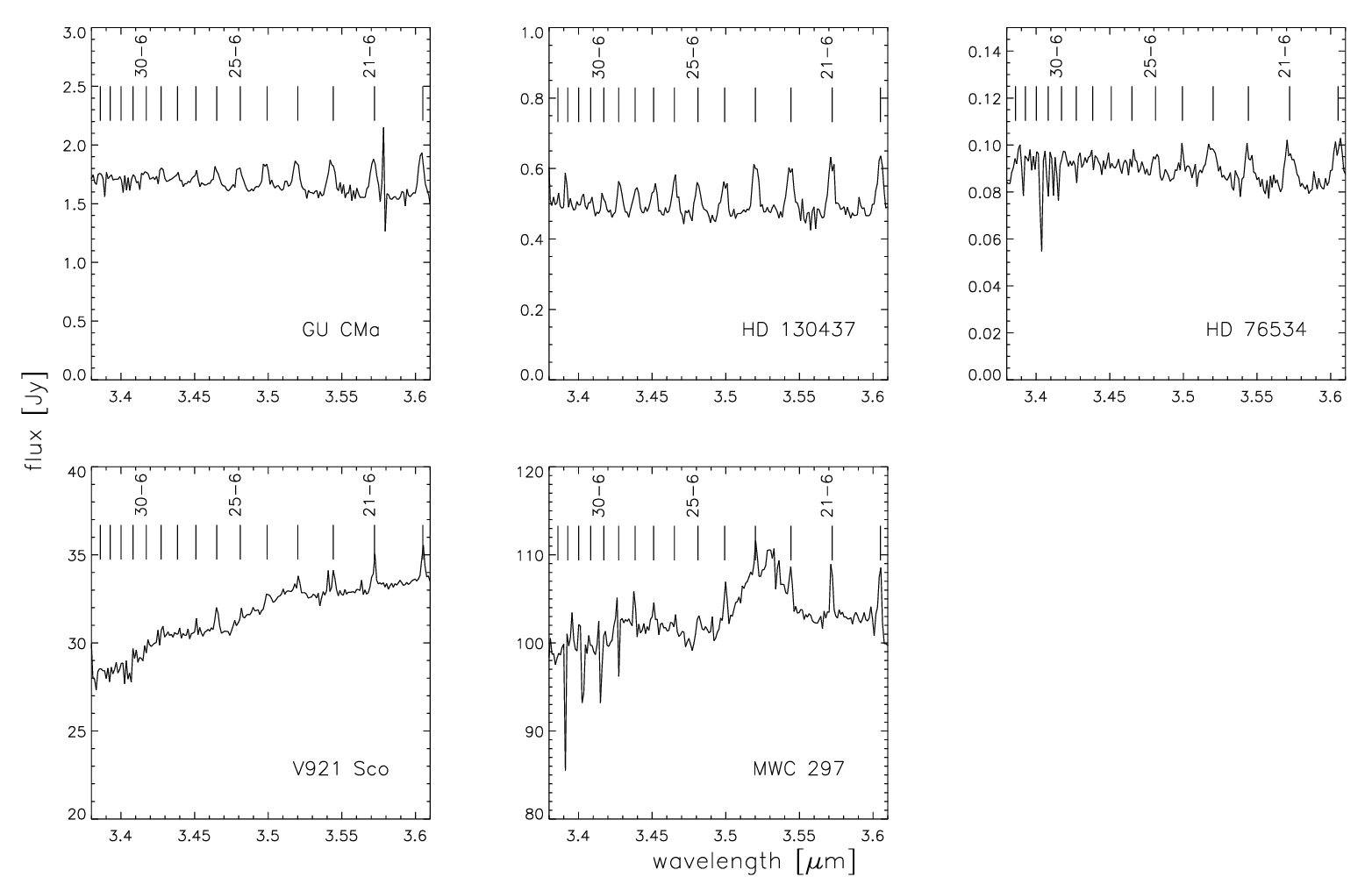}}} 
\caption{ The ISAAC 3.5 micron spectra of GU~CMa, HD~76534, HD~130437,
  V921~Sco and MWC~297. The Humphreys series of the H{\sc i} atom are
  indicated with vertical lines. The spectra in the top panels are
  clearly different from the spectra in the bottom panels. The first 3
  targets are most likely normal Be stars, with the 3~micron continuum
  flux emanating from the stellar photosphere. The latter two are
  early-type accretion-disk systems. The near-IR continuum is due to hot
  circumstellar dust.}
\label{humplot.ps}
\end{figure*}

Five of the sample stars (8\%) show the Humphreys series of the H{\sc i}
atom in emission: GU~CMa, HD~76534, HD~130437, V921~Sco and MWC~297
\citep[for the latter, see also][]{terada}. These objects are all
early-type stars (O8--B2). The SED of
GU~CMa, HD~76534 and HD~130437 does not show a significant near-IR
excess. This indicates that these sources are probably {\em normal} Be
stars, rather than Herbig Be stars. The SEDs of V921~Sco and MWC~297
on the other hand are dominated by thermal emission of the circumstellar
dust. The extinction in the line of sight towards these targets is
strong, $A_V = 5.1$ and $7.8$ respectively. These massive stars are
most likely surrounded by accretion disks. 
In Fig.~\ref{humplot.ps}, the 3.5~micron spectra of GU~CMa, HD~76534,
HD~130437, V921~Sco and MWC~297 are presented. The Humphreys series is
clearly visible.

The presence of the hydrogen Humphreys series in the 3~micron
spectra correlates well with the presence of the hydrogen \Pfd line:
the \pfd line is detected in the spectra of GU~CMa, HD~130437, 
V921~Sco and MWC~297. The 3.3~micron spectrum of HD~76534 is rather 
noisy. We have not detected an emission line at 3.296~$\mu$m exceeding
the $1\sigma$ level in the latter case.

\subsection{The PAH 3.3 micron feature}

\begin{figure}
\rotatebox{0}{\resizebox{3.5in}{!}{\includegraphics{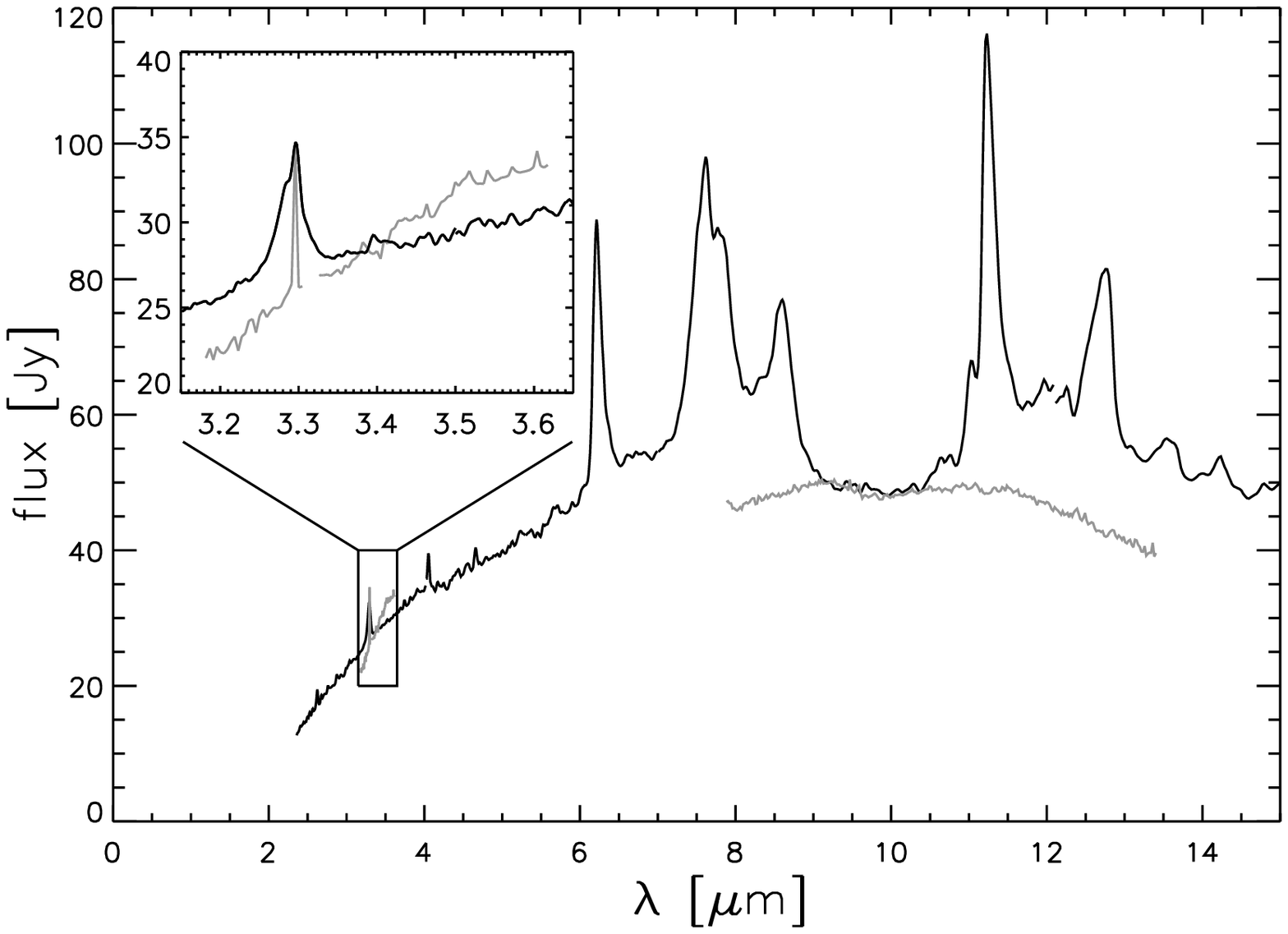}}} 
\caption{ The infrared spectrum of V921 Sco
  (CD$-$41$^\circ$11721). The black curve is the ISO spectrum
  (R=150). The grey curve around 10~micron is the long-slit TIMMI2
  spectrum (courtesy of R. van~Boekel). This spectrum is scaled to the
  IRAS 12~micron flux. For 
  clarity, a blow-up of the 3~micron wavelength region is shown in the
  inset. The grey curve in this panel is the ISAAC spectrum (R=750,
  not scaled); the black curve is the ISO spectrum at a resolution of
  500. The slope of the ISAAC spectrum clearly deviates from the slope
  of the ISO spectrum at 3~micron, most likely due to slit losses. The
  ISO spectrum is dominated by 
  strong PAH features. No PAH emission is detected in the ground-based
  spectra. The ISAAC spectrum shows the \Pfd line, the Humphreys
  series and possibly a 3.5~$\mu$m feature. In the high-resolution
  ISO spectrum, the \pfd line is visible on top of the PAH 3.3~$\mu$m
  feature. The TIMMI2 spectrum
  reveals that beneath the overwhelming PAH emission in the ISO 
  spectrum, a shallow silicate absorption band resides.}
\label{V921Sco_pahs.ps}
\end{figure}

The 3.3~$\mu$m feature, attributed to the streching vibrations of the
CH bond in PAH molecules, is detected in 7 Herbig sources (12\%). Four
additional sources tentatively show the feature. This suggests an
upper limit for the detection rate of 18\% in the ISAAC sample. The
emission band has previously been detected in the infrared spectra of
sample stars HD~97048, HD~100546 and HD~179218 (see
e.g. AV04). 
We report new detections for HD~34282, V599~Ori and AS~310. Tentative
detections are obtained for V346~Ori, V590~Mon, HD~101412 and
HBC~687. Also the ISAAC and ISO spectra of TY~CrA contain the
3.3~micron feature. Nevertheless, this source is most likely not 
a HAEBE star, considering the absence of H$\alpha$ emission and the
fact that the IR emission emanates from the ``TY~CrA~bar''
which is not directly associated to the object
\citep{siebenmorgen}. For HD~34282, we have previously
reported a non-detection of the PAH~3.3~micron feature in the
ISO spectrum (AV04). The upper limit on the line flux deduced from the
ISO spectrum is $28.1 \times 10^{-16}$ \Wm, which 
is consistent with the current detection in the ISAAC spectrum
($24\pm7 \times 10^{-16}$ \Wm). The detections of the PAH
feature in the ISO spectra of HD~142666, HD~142527, V921~Sco and
MWC~297 are not confirmed in the present spectra. For HD~142527 and
MWC~297, strong \pfd emission is observed, which is probably
misinterpreted as PAH emission in the --lower spectral resolution--
ISO spectra. Also in V921~Sco 
(=CD$-$42$^\circ$11721), strong \pfd emission is detected. The
measured line flux in the ISAAC spectrum is however not sufficient to
explain the measured ISO 3.3~$\mu$m line flux. Remarkably, also in
the ground-based
TIMMI2\footnote{http://www.ls.eso.org/lasilla/sciops/3p6/timmi/} slit
spectrum of V921~Sco in the 
10~micron wavelength region, no PAH features are detected
(Fig.~\ref{V921Sco_pahs.ps}). This may be surprizing, since the ISO
spectrum of this source is dominated by strong 
PAH emission. The explanation is found in the extent of the
emission region. Because of the
large aperture size ($14\arcsec \times 20\arcsec$), the ISO-SWS
instrument has picked up PAH emission from the wider
surroundings of the targets. Slit spectroscopy only selects a much
narrower region around the central source.
For V921~Sco, this remark is in agreement
with the result of AV04. Table~9 in that paper reveals that the
luminosity of the observed PAH emission in this source roughly
corresponds to what one may expect from a tenuous halo. No emission
features are detected in the 3.3~micron region of the ISAAC spectrum
of HD~142666, possibly indicating that also in this case, the PAH
emission is extended.

\begin{figure}
\rotatebox{0}{\resizebox{3.5in}{!}{\includegraphics{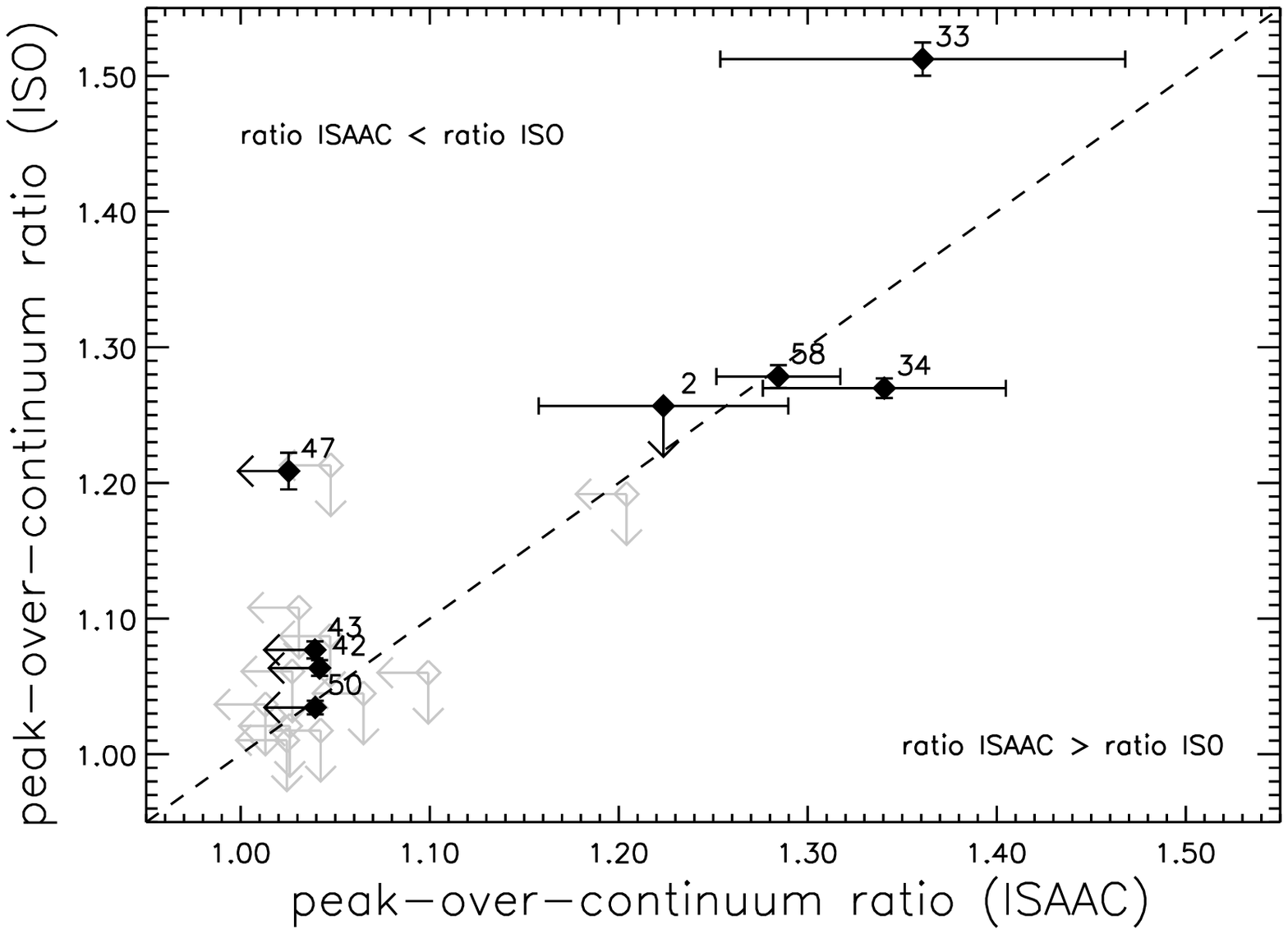}}} 
\caption{ Comparison of the peak-over-continuum ratio of the
  PAH~3.3~$\mu$m feature, measured in the ISO and ISAAC
  spectrum. Filled symbols indicate 
  targets for which we have detected the feature in at least one of the
  infrared spectra. The numbers refer to the target numbers listed in
  Table~\ref{presoffeat}. Upper limits are indicated by arrows. The dashed
  line represents ratio ISAAC = ratio ISO. }
\label{PCPAHISOvsISAAC.ps}
\end{figure}

In Fig.~\ref{PCPAHISOvsISAAC.ps} we have compared the
peak-over-continuum ratio of the PAH~3.3~micron feature measured in
the ISO and ISAAC spectrum. The measurements/upper limits of HD~34282
(\#2), HD~100546 (\#34) and HD~179218 (\#58) agree within the
error bars. For HD~142527 (\#43) and MWC~297 (\#50), strong \Pfd emission
led to a false PAH~3.3~$\mu$m detection in the ISO spectrum. In the
previous paragraph we argued that the PAH emission in HD~142666 (\#42)
and V921~Sco (\#47) is likely extended, and therefore not detected in
the ISAAC spectrum. The ISO spectra of HD~97048 show
significantly stronger PAH 3.3, NANO 3.4 and NANO 3.5~$\mu$m 
features than the ISAAC spectrum. The relative ratios between the ISO
and ISAAC line fluxes are 2.9$\pm$0.7, 2.4$\pm$0.2 and 1.65$\pm$0.02
respectively. The ISO aperture is many times larger than the slit
width of ISAAC. The PAH and nanodiamond emission emanates from the
circumstellar disk's surface. The estimated extent is of the order of
a few tens of an arcsecond \citep{vanboekel04,habart,habart04a}. This
extent, smeared out by the atmospheric seeing, could lead to a loss of
PAH and NANO emission in the ISAAC spectrum, relative to the ISO
spectrum. The underlying continuum flux, which originates from the
very inner parts of the disk, is less sensitive to this effect. This
potentially explains the weakness of the features in the ISAAC
spectrum. Alternatively, the PAH and NANO features might display
temporal variability. This variability has not been noted in the five
ISO 3~micron spectra which are available for this source,
however. These ISO observations span about one year and a half
(AV04).

The ISAAC detection rate of the PAH 3.3~$\mu$m feature in the group~I 
sources is 36\% (50\% including tentative detections). The rate in
group~II is much lower (0$-$3\% respectively). A PAH
3.3~$\mu$m feature was detected in the spectrum of the embedded source
V599~Ori (17\%). In the
computation of these numbers, we have excluded GU~CMa, HD~76534,
HD~130437 and TY~CrA, which are most likely no Herbig stars. The high
and very low detection 
rate in respectively group~I and II confirms the main conclusion of
AV04. The PAH features emanate from the surface of a flared
circumstellar disk. Only group~I sources have a flared disk and are
expected to display strong PAH features \citep[][AV04]{habart}. In
Table~\ref{detectionrates} we compare the detection rates of the PAH
3.3~micron feature in the ISO sample to the rates in the present ISAAC
sample. The ISO sample contains 16 group~I, 22 group~II and 7 embedded
sources\footnote{We have classified T~CrA as group~II member and
V921~Sco as embedded source in stead of in group~I (as was done by
AV04).}. The combined ISO+ISAAC sample consists 
of 25 group~I, 42 group~II, 11 embedded and 4 unclassified
targets, in total 82 sources. A number of targets are present in
both the ISO and ISAAC sample. The sources with a PAH detection in the
ISO spectrum but not in the ISAAC spectrum, or vice versa, were all
discussed in the previous paragraphs. We have consistently included
these results in the computation of the detection rates. Up to half of
the sources classified in group~I display the PAH~3.3~$\mu$m feature,
while it is detected in only a small number of group~II and embedded
objects.

\begin{table}
\caption{ The detection rates of the PAH 3.3~$\mu$m feature in the ISO
  and ISAAC samples. T stands for the total of classified sample
  sources, I and II refer to the Meeus groups and E represents the
  embedded sources. The last
  row gives the detection rates in the combined ISO+ISAAC sample. When
  two percentages are given, the second refers to the detection rate
  including tentative detections. The detection rate in group~I is
  significantly higher than in group~II (confidence level
  $>$~99.5\%).
\label{detectionrates}}
\begin{center}
\begin{tabular}{r|cccc}
       & T       & I       & II       & E       \\
\hline	                   	      	        
ISO    &22$-$29\%&44$-$56\%& 14\%     & 0$-$14\%     \\
ISAAC  &12$-$18\%&36$-$50\%& 0$-$3\%  & 17\%    \\
total  &14$-$21\%&32$-$48\%& 5$-$7\%  & 9\%     \\
\end{tabular}
\end{center}
\end{table}

According to AV04, the PAH 3.3~micron
feature is the least detected PAH feature in the 3-11~micron region in
the ISO sample. It is present in only 44\% of the cases where the
6.2~micron feature is detected. The latter is the most detected PAH
feature in the ISO study. Furthermore, the PAH 6.2 feature is {\em
always} detected when the PAH 3.3~$\mu$m feature is
present\footnote{Except for HD~142527, a PAH 3.3~$\mu$m detection
which was however not confirmed by the ISAAC spectrum.}. Using
Bayes's theorem\footnote{$P(B) = \frac{P(B|A)~\cdot~P(A)}{P(A|B)}$ with
A and B stochastic events. In this case A and B represent the detection
of the 3.3 and the 6.2~$\mu$m feature in the spectrum, respectively.},
one can estimate the hypothetical detection 
probability of the 6.2~micron feature in the total sample. We assume
that the detection probability of the 3.3~micron feature is equal to
the relative detection rate in the total sample. The conditional
probabilities --the relative detection rate of one feature, given the
presence of the other-- are taken from the ISO study (AV04). For
group~I, the hypothetical PAH 6.2~micron detection probability is
73\%$-$100\%, while it is 11\%$-$16\% for group~II and 20\% for the
embedded sources respectively. Hypothetically, three quarters to all
flared disk sources are PAH emitters. Most of the PAH detections in
group~II and in the embedded sources, as well as a fraction of the
group~I detections, are likely due to extended emission from a
circumsystemic nebula. 

\subsection{The nanodiamond features \label{sectnano}}

It is clear that our ISAAC sample does not contain any targets with a
high 3.53~$\mu$m feature/continuum ratio except HD~97048.
This in itself is an important result. The detection of a
strong 3.53~$\mu$m feature in the near-IR spectra of Herbig stars
is thus limited to Elias~3$-$1, HD~97048 and MWC~297. This leads to a
detection rate of 3/82 = 4\% for the entire ISO+ISAAC sample. The
3.43~$\mu$m feature is even more rare, since it has been detected
in Elias~3$-$1 and HD~97048 only (2\% of the sample).

We confirm the detection of the 3.53~micron feature in
MWC~297 \citep[][AV04]{terada}. The shape of this complex band
resembles that of the 3.53~$\mu$m feature in HD~97048, in the
sense that the red wing is steeper than the blue wing and the peak
position is around 3.529$\pm$0.002~$\mu$m. It is therefore likely that
these features have a common carrier, i.e. (nano)diamonds. The line
flux ratio of the 3.43 and 3.53~$\mu$m features is 0.82 (ISO) for
Elias~3$-$1 and 0.41 (ISAAC; ISO: 0.59) for HD~97048. The upper limit
to this ratio is 0.39 (ISO; ISAAC: 2.0) for MWC~297. This indicates
that the NANO 3.43~$\mu$m is, if present, at least weaker than in the
two other (nano)diamond sources.

A few more sample targets display a feature around 3.5~$\mu$m.
In the Appendix, we argue that most of these features are
artefacts. Only for V921~Sco, HD~163296 and T~CrA, a potentially real
feature is detected. These detections are by far not sure, however. In
the following, we refer to these sources as the candidate nanodiamond
emitters.

\section{What makes the nanodiamond emitting sources so special?}

It is clear that very few targets display nanodiamond features in
their 3~micron spectra. The main question which is addressed in this
paper concerns the origin of these features. In this section, we will
show that the 3.43 and 3.53~$\mu$m emitting sources are no outliers
in the group of Herbig stars as a whole (1), but also that these
targets have very different stellar and circumstellar parameters
compared to each other (2).

\subsection{Nanodiamonds and polycyclic aromatic hydrocarbons}

\begin{figure}
\begin{center}
\resizebox{3.5in}{!}{\rotatebox{0}{\includegraphics{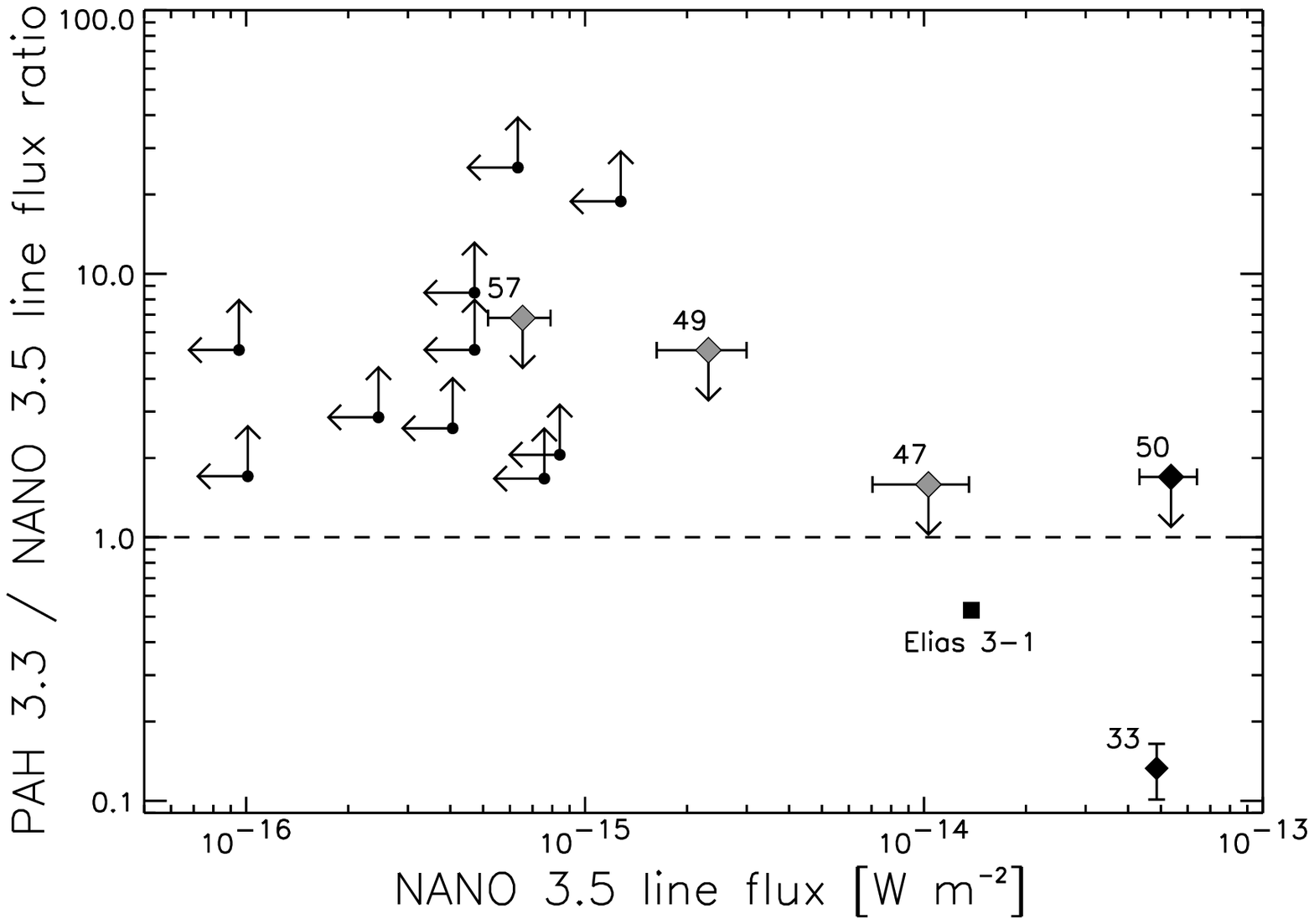}}} 
\caption{ The line flux ratio of the PAH~3.3~$\mu$m and the
  nanodiamond 3.53~$\mu$m feature versus the NANO~3.53~$\mu$m
  line flux. The dashed line indicates where both line fluxes are
  equal. The black square indicates the ISO measurements of
  Elias~3$-$1. The black diamonds
  represent HD~97048 and MWC~297. The grey diamonds are the
  nanodiamond candidates V921~Sco, HD~163296 and T~CrA. The black dots
  represent the targets with a detected
  PAH~3.3, and undetected NANO~3.53 feature. Only in HD~97048 and
  Elias~3$-$1, both features are detected. The line flux ratio of the
  PAH~3.3 and NANO~3.53 features is smaller than one in these targets,
  while the upper limits on this ratio are larger than unity in the
  other (candidate) nanodiamond emitters.}
\label{pahvsnano.ps}
\end{center}
\end{figure}

PAHs and nanodiamonds are both carbonaceous grains. The 3.3, 3.43 and
3.53 features have been attributed to the C$-$H bonds in these
structures. The emission of
both species has been shown to be spatially extended and linked to the 
surface of the circumstellar disk
\citep[][AV04]{vanboekel04,habart,habart04a}. Fig.~\ref{pahvsnano.ps}
displays the PAH 3.3~$\mu$m line flux versus the nanodiamond
3.53~$\mu$m line flux. The line fluxes determined for Elias~3$-$1 are
included as well. The latter are based on the ISO spectrum of this
source, rebinned to a resolution of 500. For the sake of clarity, we
left out the upper limits when both features were undetected. 
Although a correlation between PAHs and nanodiamonds may be expected
based on their common chemical root, the PAH
3.3~$\mu$m feature is not detected in MWC~297 and the three tentative
nanodiamond emitters V921~Sco, HD~163296 and T~CrA. Furthermore, the
PAH 3.3~$\mu$m feature is obviously much more common than the
nanodiamond features in the group of HAEBEs. There does not seem to be
a connection between them. 
However, it should be noted that the line flux ratio of 
PAH~3.3~$\mu$m and NANO~3.53~$\mu$m in HD~97048 and Elias~3$-$1 --the
only two sources where both features are detected-- is
significantly smaller than 1 (0.13$\pm$0.03 and 0.5$\pm$0.1,
respectively). The upper limit for this 
ratio in the other four (1 + 3 candidate) nanodiamond targets is
larger than unity. Hence it may be that the PAH~3.3~$\mu$m feature is
present, but too weak to 
be detected in the latter sources. If it is always true that the line
flux ratio of PAH~3.3~$\mu$m and NANO~3.53~$\mu$m is smaller than one,
the upper limits on the NANO~3.5 line flux in the other stars with
detected PAH emission indicate that no nanodiamond feature
can be present; the lower limits on the line flux ratios are all
above one in these targets. In conclusion, the nanodiamond
feature at 3.53~$\mu$m appears to be stronger than the PAH~3.3~$\mu$m
feature. In the other sense, the presence of a 
PAH~3.3~$\mu$m feature does not imply the presence/absence of a
nanodiamond feature.

\subsection{Appearance of the disk}

As mentioned before, the (nano)diamond features are linked to the
circumstellar disk. Here we investigate whether the disks of MWC~297,
Elias~3$-$1 and HD~97048 are in any way different from other HAEBE
disks. We have summarized some overall disk parameters of the
(candidate) nanodiamond sources in Table~\ref{diskpar}. As a
reference, we list the median value of each parameter in each
group. The median values of $L_\mathrm{exc}/L_\star$ and the spectral
index (a proxy for dust particle size) are not significantly different
in sources with nanodiamond features as opposed to the sources without.

\begin{table}
\begin{minipage}[t]{\columnwidth}
\tabcolsep2pt
\caption{ Some disk parameters of the (candidate) nanodiamond
  sources. The Meeus group indicates the overall geometry of the
  disk. The ratio of the total IR excess to the stellar luminosity,
  $L_\mathrm{exc}/L_\star$, indicates the fraction of stellar flux
  that is captured and reprocessed by the disk. For passive disks, this
  number represents the fraction of the sky covered by the disk, as
  seen from the central star. The ratio for T~CrA is
  likely overestimated due to strong nearby IR emission, not directly
  linked to the system. The parameter $n$ is the spectral index in the
  450~$\mu$m$-$1.3~mm range, defined by $\lambda F_\lambda \propto
  \lambda^{-n}$.
\label{diskpar}}
\renewcommand{\footnoterule}{}
\begin{center}
\begin{tabular}{llccc}
\hline\hline
\# & Target  & Group & $L_\mathrm{exc}/L_\star$ & $n$ \\
\hline	                   	      	        
   & Elias~3$-$1    & I & 0.5$-$1.3\footnote{For Elias~3$-$1 we used a stellar
  luminosity of 20-50~$L_\odot$ (see Table~\ref{stelpar}).} & 3.23$\pm$0.03  \\
33 & HD~97048       & I & 0.40 & $>$2.9  \\
50 & MWC~297        & E & 0.06 & 3.1$\pm$0.2  \\
\hline		           
47 & V921~Sco       & E & 0.11 & 4.1$\pm$0.3  \\
49 & HD~163296      & II& 0.34 & 2.9$\pm$0.1  \\
57 & T~CrA          &II?& 3.8  & ?  \\
\hline		        
   & Median & I & 0.50 & 3.9  \\
   & Median & II& 0.44 & 3.0  \\
   & Median & E & 0.63 & 4.1  \\
\hline
\end{tabular}
\end{center}
\end{minipage}
\end{table}

\subsection{Stellar characteristics}

MWC~297, Elias~3$-$1, HD~97048 and the three candidate nanodiamond
stars have very different stellar parameters. We have summarized
their spectral type, estimated age and stellar luminosity in
Table~\ref{stelpar}. The median parameters for the total sample are
again given as a reference. Nanodiamond features are found in sources
with a wide range of spectral types (B1.5--F0). Although the
nanodiamond sources might be a bit older ($>$~1~Myr) than
typical for Herbig stars, there exists no strong correlation with age;
there are many examples of Herbig stars with similar ages in our
sample which do not show nanodiamond features of comparable strength.

The absolute strength of the nanodiamond features (i.e., the
nanodiamond luminosity, included in Table~\ref{stelpar}) appears to be
linked to the spectral 
type (and hence $T_\mathrm{eff}$ and luminosity) of the central
star. This weak correlation appears to be the only one which can be
derived. The nanodiamonds in the circumstellar disk must be heated to
a certain temperature to be able to radiate in the near-IR. Hotter
stars heat larger volumes of circumstellar dust. The extent of the
nanodiamond emission region is therefore expected to be larger around
earlier-type stars. This is in agreement with the observed correlation.

\begin{table}
\tabcolsep2pt
\caption{ Some stellar parameters of the (candidate) nanodiamond
  targets, and the observed nanodiamond 3.53~$\mu$m luminosity
  $L_{3.5}$. The median value of each stelllar parameter in the total
  sample is given as a reference. Values taken from \citet{houk,
  finkenzeller84, shore, vandenancker98, mora01,fuente01, habart03,
  ackesubmm, hamaguchi05}, AV04 and references therein.
\label{stelpar}}
\begin{center}
\begin{tabular}{llcccc}
\hline\hline
\# & Target  & Sp. Type & log Age/yr & $\log L_\star / L_\odot$ & $L_{3.5} / L_\odot$\\
\hline	                   	      	        
   & Elias~3$-$1 & A6:e      &6.3$-$6.9& 1.3$-$1.7& $1 \times 10^{-2}$ \\
33 & HD~97048    & B9.5Ve+sh &$>$6.3   & 1.9      & $5 \times 10^{-2}$ \\
50 & MWC~297     & B1.5Ve    &5$-$6    & 4.2      & $1 \times 10^{-1}$ \\
\hline					            
47 & V921~Sco    & B0IVep    &5$-$6    & 4.9      & $8 \times 10^{-2}$ \\
49 & HD~163296   & A3Ve      & 6.6     & 1.6      & $1 \times 10^{-3}$ \\
57 & T~CrA       & F0e       & ?       & 0.8      & $3 \times 10^{-4}$ \\
\hline					            
   & Median      & A2        & 6.0     & 1.7      &  \\
\hline					            
\end{tabular}
\end{center}
\end{table}

One of the current hypotheses on the presence of the nanodiamonds in
our own solar system suggests that they were formed in a supernova
outflow, and injected into the system \citep[idea originally put
forward by][see Sect.~\ref{discussion}]{clayton89}. 
Under this hypothesis, the presolar nebula is enriched in matter from
the supernova outflow. The parent star of a disk which contains
supernova-injected diamonds must therefore be polluted by supernova
material as well. Supernova outflows harbor $r$- and $p$-process
elements and specific isotopes of 
lighter elements. Enrichment of the stellar photosphere by such an
outflow is expected to alter the photospheric abundance pattern of the
star significantly. An abundance analysis of the nanodiamond
sources opposed to stars with comparable stellar parameters might
clarify whether or not supernovae are indispensable for the formation
of (nano)diamonds in their circumstellar disks. For HD~97048, two
high-quality optical spectra are available to us: a FEROS
3900$-$9100\r{A} spectrum with a signal-to-noise ratio (S/N) of 115,
and a UVES 4800$-$6800\r{A} spectrum with a S/N of 460. The latter
spectrum is compiled from 12 UVES spectra, obtained in a time span of
a few days. The spectra have been previously published in
\citet{ackelamboo} and \citet{ackeoi2}, respectively. Based on the
Vienna Atomic Line Database \citep[VALD,][]{kupka99},
we have looked for the strongest 
absorption lines of elements with atomic number larger than 30
(Zn). At none of the VALD wavelengths, an absorption line is
detected. We have deduced upper limits on the equivalent width of the
strongest lines and hence on the abundances of four heavy
elements. The abundances of Y, Zr and Ba in HD~97048 are less than a
few times the solar abundance. Strontium is even depleted by at
least a factor 10 compared to our sun. We conclude that there is no
observational evidence for a photosphere enriched with heavy elements
in the nanodiamond source HD~97048, and hence no evidence for the
occurance of a recent supernova event in the target's proximity.

\subsection{Disk and stellar activity}

\begin{figure}
\begin{center}
\resizebox{3.5in}{!}{\rotatebox{0}{\includegraphics{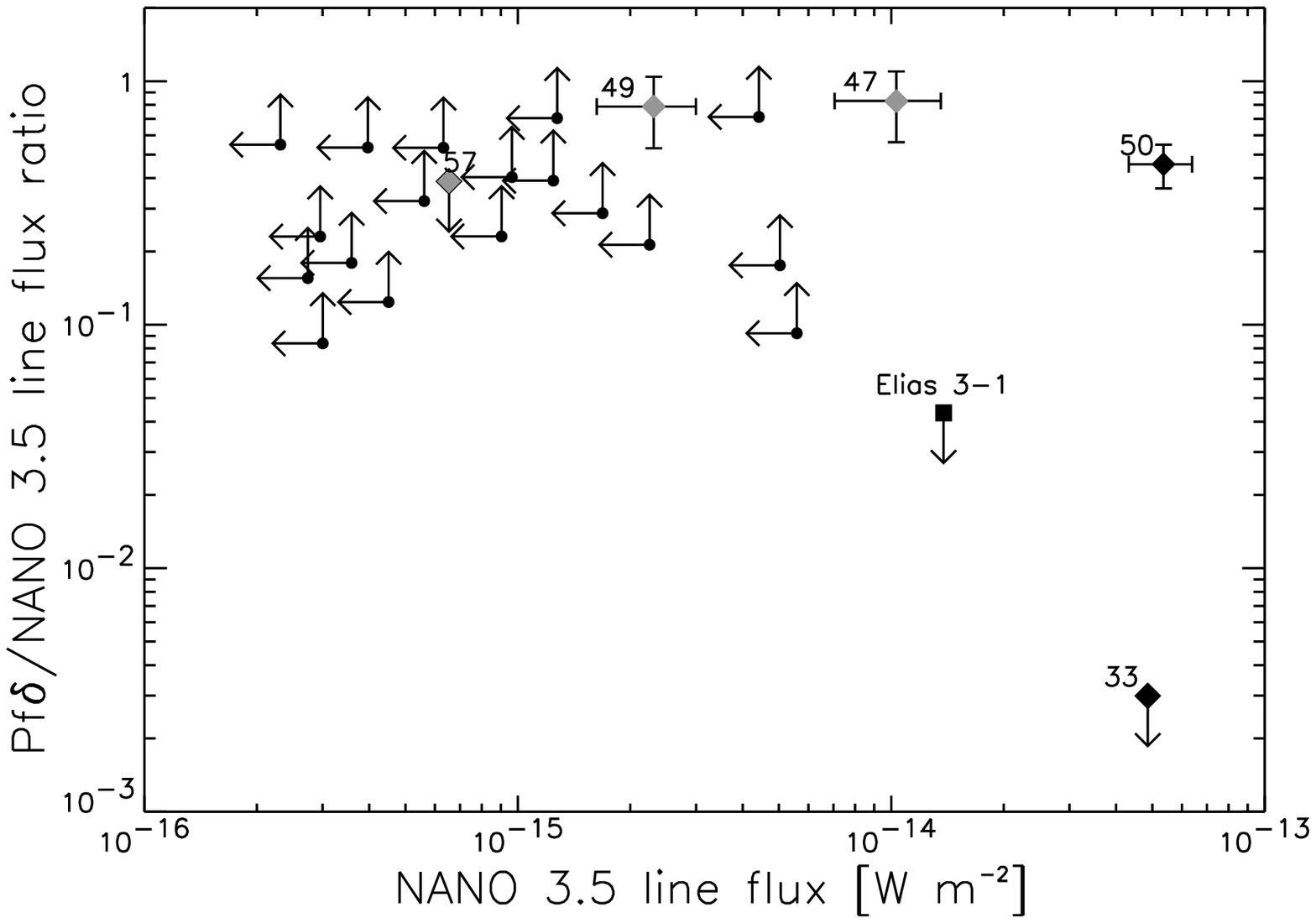}}} 
\caption{ The line flux ratio of \pfd and NANO~3.53~$\mu$m versus the
  NANO~3.53~$\mu$m line flux. Legend see Fig.~\ref{pahvsnano.ps}.}
\label{pfdvsnano.ps}
\end{center}
\end{figure}

In Fig.~\ref{pfdvsnano.ps}, the line flux ratio of the \pfd line and
the nanodiamond 3.53~$\mu$m feature is plotted versus
the NANO 3.53~$\mu$m line flux. Group~I members HD~97048 and
Elias~3$-$1 have no detected \pfd 
line. The massive young stellar object MWC~297 does display both
features. We have indicated the measurements for the tentative
nanodiamond emitters (V921~Sco, HD~163296 and T~CrA) with grey
plotting symbols. Also for these three targets, there is no
uniformity. The early-type embedded source V921~Sco and group~II
member HD~163296 display the \pfd line, while the UXOR object T~CrA
does not. 

X-ray emission in stars with convective envelopes, like the lower-mass
T~Tauri stars, is often explained as being due to the presence of a
magnetic field. The matter is less clear for the earlier-type Herbig
stars. Although these stars are believed to have radiative envelopes
and thus cannot drive a solar-like magnetic dynamo, the X-ray
luminosity of Herbig stars is nevertheless explained in the light of
magnetic effects. Such a magnetic field can be fossile
\citep[e.g.][]{skinner04,hamaguchi05}, or produced differently than in
our Sun \citep[e.g. the differential rotation dynamo
model,][]{tout95}. Hard and soft X-ray spectra are observed in the 
group of HAEBE stars. Generally, the first kind is attributed to hot
plasma in a stellar corona or the presence of a late-type companion,
the second to accretion events onto the 
photosphere. In O-type stars, energetic wind shocks can produce X-ray
emission. It appears that Herbig winds are not powerful enough to
produce the same effect \citep{skinner04,hamaguchi05}. 

\begin{figure}
\begin{center}
\resizebox{3.5in}{!}{\rotatebox{0}{\includegraphics{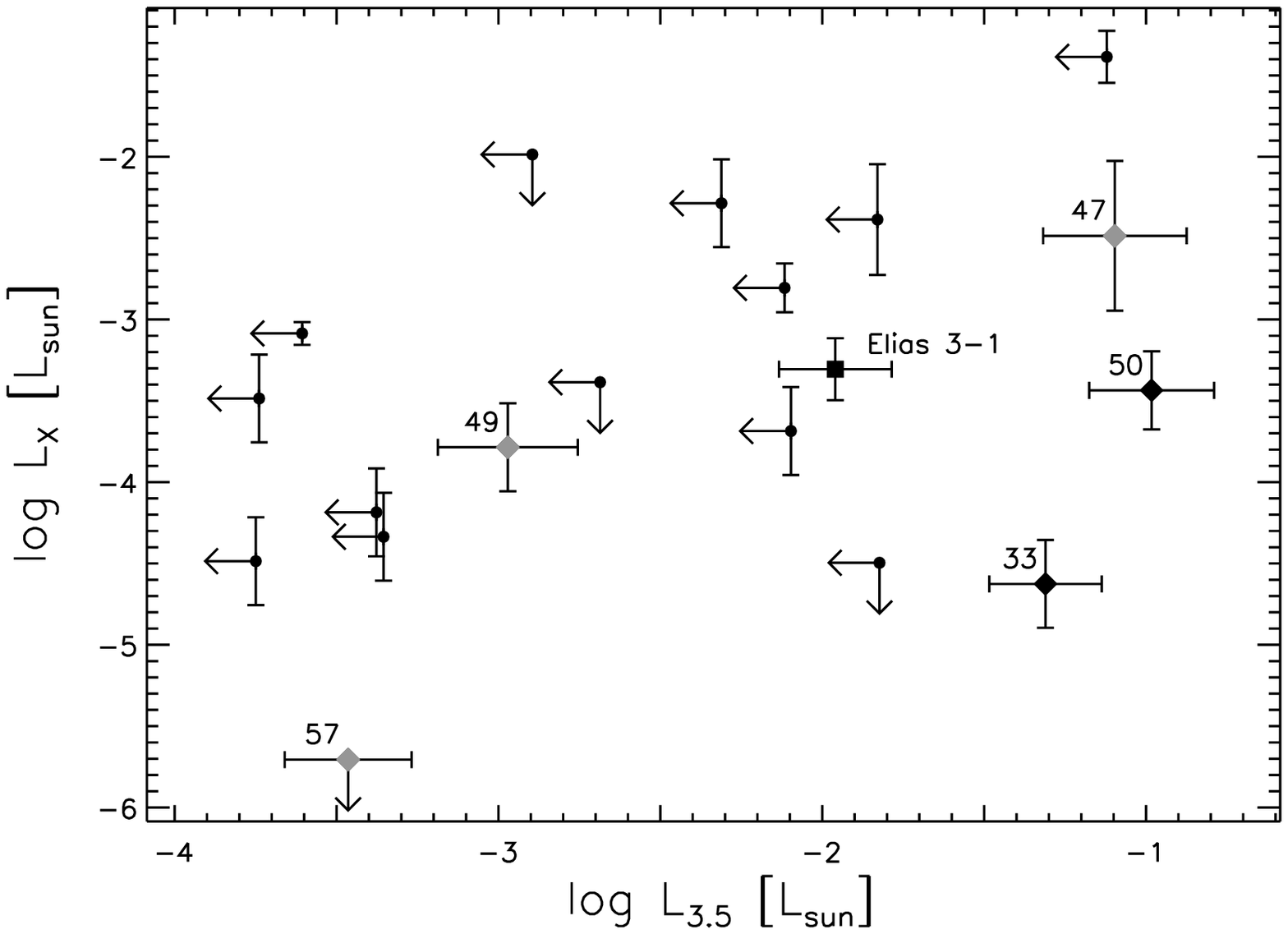}}} 
\caption{ The X-ray luminosity $L_\mathrm{X}$ versus the
  NANO~3.53~$\mu$m line luminosity $L_{3.5}$. The latter quantity is
  the line 
  flux rescaled by a factor $4\pi\ d^2$, with $d$ the distance to the
  source as listed in \citet{skinner04}, \citet{hamaguchi05} or
  \citet{damiani06}. We apply the same distance factor to the
  3.53~$\mu$m line flux as was applied for the cited
  $L_\mathrm{X}$, to avoid systematic errors. The plotting symbols are
  the same as in 
  Fig.~\ref{pahvsnano.ps}. The (candidate) nanodiamond emitting 
  sources have X-ray luminosities which are not uncommon in the
  group of sample stars observed in the X-ray wavelength
  range.}
\label{xrayvsnano.ps}
\end{center} 
\end{figure}

Whatever
mechanism causes the X-ray luminosity in Herbig stars, it does not
seem to influence the presence/absence of the nanodiamond features. In
Fig.~\ref{xrayvsnano.ps}, we have plotted the X-ray luminosity versus
the NANO~3.5 line luminosity. The X-ray luminosities are taken from
\citet{skinner04}, \citet{hamaguchi05} and \citet{damiani06}. It
appears that, in this small sample of observed Herbig stars, the
(candidate) nanodiamond emitters have typical X-ray luminosities. When
rescaling the NANO~3.5 line flux to the appropriate distance to the
source, MWC~297 appears to be the most luminous nanodiamond target
($L_{3.5} = 0.10~L_\odot$). It is a
rather strong X-ray source, although the X-rays likely emanate
from nearby lower-mass stars \citep[we follow the suggestion that
5\% or less of the total X-ray luminosity around MWC~297 emanates
from the B1.5 star itself,][]{vink05b,damiani06}. Elias~3$-$1 has a 
comparable X-ray luminosity, but a 3.53~$\mu$m luminosity which is ten
times smaller. HD~97048, the third confirmed nanodiamond emitter, has
an intermediate nanodiamond luminosity ($L_{3.5} = 0.05~L_\odot$), but
an X-ray luminosity 15 
times more modest than the previous two targets. Recently, Elias~3$-$1
has been identified as a binary \citep[with a
separation of 9~AU at a distance of 160~pc,][]{smith05}. The companion
is a lower-mass star, possibly with a convective envelope. The
latter may explain the rather high X-ray luminosity attributed to
Elias~3$-$1.

\vspace*{0.3cm}

The analysis presented in the previous sections suggests that the few
sources 
which display nanodiamond features are not exceptional compared to the
bulk of Herbig stars. Furthermore, these targets do not appear to have
obvious stellar or circumstellar characteristics in common.

\section{Conclusions and Discussion \label{discussion}}

We summarize the two main conclusions of the present paper.

\begin{itemize}
\item 
  Our survey of 3~micron spectra of HAEBE stars has not revealed a new
  spectacular nanodiamond source like HD~97048 or Elias~3$-$1. This
  negative result increases the rarity of these
  near-IR bands in the group of Herbig stars. Within the current
  sample, only 4\% of the targets are surrounded by detected nanodiamonds. 
  We find
  a large range in nanodiamond luminosity, which spans 1$-$2 orders of
  magnitude. No clear correlation is found between the presence of the
  nanodiamond features and the disk and stellar
  parameters, although a weak link between nanodiamond luminosity and
  stellar luminosity may be present. Furthermore, there does not appear to
  be a specific disk/star property which the nanodiamond sources have
  in common. We have shown that our findings are inconclusive with
  respect to the influence of the surroundings of the targets. We do
  not find any evidence for a correlation of nanodiamond 
  emitters with the 
  locations of known recent supernova remnants, nor do we find evidence
  for an enrichment in heavy elements of the photosphere of the system
  with the most prominent nano-diamond features, HD~97048.
\item Most to all Herbig stars with a flared disk are PAH
  emitters. This conclusion once again confirms the suggestion of
  \citet{meeus01} that PAH molecules can only be excited in systems
  where they can capture direct stellar UV photons. In the case of
  self-shadowed disks, the puffed-up inner rim casts a shadow over the
  outer disk, preventing the PAH molecules from being heated by the
  central star. Our findings reinforce the results of AV04
  and \citet{habart}. The latter studies were based on ISO data. Due
  to the large aperture sizes of the ISO instruments, it is difficult
  to differentiate between PAH emission emanating from the disk
  surface, and contamination from the wider surroundings of the
  target. The present ISAAC data set, which clearly confirms the
  correlation, is considerably more reliable in this sense: both the
  narrow slit and the chop-nod technique, which allows for an
  efficient subtraction of constant background emission, reduce the
  possibility of confusion with extended PAH emission drastically.
\end{itemize}

Laboratory experiments have been instrumental in the
identification of the carriers of the 3~micron features
\citep[e.g.][]{guillois,sheu02,jones04,mutschke04}. Currently, 
the astronomical spectra can be reproduced from the absorption spectra
of films of hydrogen-terminated diamonds \citep{sheu02}. The spectral
features in the 3~micron range are due to CH-bond stretches (as is the
PAH~3.3~$\mu$m feature). The prominent peaks
around 3.43 and 3.53~$\mu$m observed in HAEBE spectra can only be
reproduced when the diamonds are sufficiently large. Typical sizes for
the diamonds in HAEBE stars ($\sim$50-100~nm) are 2 orders of magnitude
larger than the nanodiamonds found in our solar system
\citep[e.g.][]{jones04}. The spectral 
features hence indicate the presence of large, hot
hydrogen-terminated diamonds.

The question remains whether there are indeed no diamonds in the
majority of disks. It may be possible that nanodiamonds of small sizes
are present, but remain undetected, because their spectral 
signature is less pronounced \citep[e.g.][]{sheu02}. Also diamonds
which are not hydrogen-terminated are not detectable in the 3~micron
region. Diamonds that are stripped from hydrogen atoms are hence
overlooked by our survey. A third possibility why (nano)diamonds can
remain undetected is their exact location |and hence temperature| in
the disk. Cool hydrogen-terminated diamonds will not radiate
significantly in the near-IR. The absence of the 3~micron features in
the spectrum of a HAEBE system hence does not necessarily mean that no
(nano)diamonds are present.

\subsection*{In-situ formation or extraneous origin?}

Diamond was the first presolar mineral to be identified in meteorites
\citep{lewis87} and it has the highest relative abundance among
carbonaceous grains. The diamond has been classified as presolar based
on the presence of a significant overabundance of very heavy (Xe-H)
{\em and} very light (Xe-L) Xe isotopes (together Xe-HL). The
overabundance of this noble-gas carrier is expected to be produced by
$r$- and $p$-process nucleosynthesis in supernovae. The link between
the presence of nanodiamonds and the xenon overabundance led to the
hypothesis that the diamonds are formed in, and injected into the
system by a supernova outflow \citep{clayton89}. Isotope anomalies on
other heavy elements such as Te and Pd also point to supernova
nucleosynthesis \citep{richter98, maas01}. 

Recent analysis of $^{60}$Fe isotopes in chondrites
has clearly demonstrated that a nearby type~II supernova has
contributed material to the natal cloud from which our own solar
system formed \citep{mostefaoui05,tachibana06}. 
This process may have been important for planet formation, as the
the radioactive decay of $^{60}$Fe into $^{60}$Ni was an important
heat source for the early planetary melting and differentiation
and keeping asteroids thermally active for much longer than would be
possible from the decay of $^{26}$Al alone.

Infrared spectroscopy has
shown that solar-system diamonds contain N and O, probably in chemical
functional groups on their surfaces
\citep{lewis89,mutschke95,andersen98,jones04}. However, the similarity
of C- and N-isotopes of these diamonds and the solar system as a whole
supports the idea that not all solar-system diamonds originate from
supernovae and that the supernova contribution to the diamonds in our
own solar system is probably not very large. It thus remains unclear
how the majority of the solar-system diamonds has formed.

Although advocated by several recent studies
\citep{kouchi05,binette05}, the presence of 
nanodiamonds in the interstellar medium remains controversial.
The absence of observable 3.43 and 3.53~$\mu$m features in the ISM
suggests that hydrogenated hydrocarbons cannot be more abundant than
$\approx$~0.1 ppm in the interstellar medium \citep{tielens87}.
\citet{vankerckhoven} showed that 
in HD~97048 and Elias~3$-$1, 1 nanodiamond part per billion relative
to hydrogen is required to reproduce the 3~micron spectra observed
by ISO, making an ISM origin of nanodiamonds in principle a viable
option. This hypothesis has problems in explaining the paucity
of detections of the 3.43 and 3.53~$\mu$m feature in Herbig stars
within our sample, however.

We found no evidence for the presence of a
supernova remnant (SNR) in the proximity of any of our sample
sources. We made use of the catalog of galactic SNRs
published by \citet{green04}. Only R~Mon and
V590~Mon are potentially located close to a SNR; the Monoceros Nebula
(G205.5$+$0.5). These sources display no detected nanodiamond features,
however. We note that the absence of a SNR in the vicinity of our
sample stars in the above-mentioned catalog does not necessarily imply
that no supernova has actually occured. Most of the sample targets are
located in massive-star-forming regions. It may very well be that the
most massive members in such a region have recently gone off as a
supernova. In a relatively crowded star forming region, the ejecta of
such an event likely affect a number of surrounding young stellar
objects or molecular clouds. In this scenario, the presence of
diamonds in the circumstellar 
disk is expected to be common to all disk systems in the vicinity of the
supernova. Unfortunately, we have no spectra of disk sources close to
Elias~3$-$1, HD~97048 or MWC~297 to check this hypothesis.

The parent star of a disk which contains
supernova-injected diamonds must be be polluted by supernova
material as well, under the supernova hypothesis. Enrichment of the
stellar photosphere by a supernova outflow is expected to alter the
photospheric abundance pattern of the 
star significantly. An abundance analysis of the strongest diamond 
source HD~97048 has produced upper limits on the abundances of a few
heavy elements. The Sr abundance is at least one order of magnitude
less than the solar abundance of this element. No evidence supporting
the supernova hypothesis is found.

Except for HD~97048 and Elias~3$-$1, both nanodiamond features have
also been observed in the post-AGB binary HR~4049
\citep{geballe89}. The post-AGB phase is very short
($\sim$10,000~yr) compared to the lifetime of an intermediate-mass
pre-main-sequence star. 
Due to the short timescales, it is 
unlikely that a nearby supernova is the cause of the presence of
nanodiamonds in this system. However, the oxygen isotopes in the
circumbinary disk of HR~4049
display peculiar behaviour: the relative $^{17}$O and $^{18}$O abundances
are two orders of magnitude larger than the surface abundances in
other evolved stars \citep[$^{16}$O/$^{17}$O~=~8.3$\pm$2.3 and
$^{16}$O/$^{18}$O~=~6.9$\pm$0.9,][]{cami01}. These exceptional values are
most likely related to the binary nature of HR~4049, and possibly to a
nova outburst. The presence of nanodiamonds in the circumstellar
environment may be related to such an event as well. 

The extraneous origin of the circumstellar diamonds is disputed,
however. \citet{dai02nature} have stated that, if the
supernova hypothesis is true, nanodiamonds should be abundant in
solar-system comets as well. These objects are believed to have formed
further out in the early solar system and are likely more pristine
than meteorites. The authors have investigated fragile, carbon-rich IPD
particles which enter the Earth's atmosphere with
speeds in the range of cometary bodies, which suggests that the grains
originate from these objects. Nanodiamonds are absent or
strongly depleted in such IPD grains, which indicates that diamonds are
not uniformly abundant in the solar system. This may support
the hypothesis that the detected meteoritic nanodiamonds have formed
{\em in situ}, and are not of presolar origin.

\citet{goto00} has suggested that the X-ray hardness of the radiation 
field of the central star may play a decisive role in the formation of
diamond. However, our analysis has shown no obvious link between the
source's 
X-ray luminosity and presence/absence of nanodiamonds. Furthermore,
the strength of the few detected diamond features appears to be
independent from the X-ray strength of the central star as
well. 

The extraneous origin of nanodiamonds is an attractive
hypothesis, because it naturally explains the paucity of nanodiamond
sources and the seemingly very different properties of these
targets. No evidence was found, however, to substantiate the claim of 
an external triggering source in the vicinity of the targets.
In-situ formation of nanodiamonds is a viable alternative. However, it 
remains unclear why only very few disks can/could provide the specific
conditions needed to produce diamond. The disk and stellar parameters
of the diamond sources are not uncommon, nor do the targets form a
consistent group within the Herbig stars. Selection effects in terms
of the size, temperature or hydrogenation of the diamonds may
nonetheless be present. Such effects could prevent the detection of
features in the 3~micron region in spectra of the majority of
circumstellar disks.
A larger sample of systems displaying diamond emission is needed
to distinguish between in-situ formation and an extraneous origin of
the (nano)diamonds in Herbig Ae/Be stars and in our own solar system.

\section*{Appendix}

Apart from HD~97048 and MWC~297, a few other targets display a feature
in the 3.5~micron region. Due to their low feature/continuum ratio,
however, it is difficult to discern them from data reduction
artefacts. In this Appendix we will try to clarify which features are
astrophysical, and which are not. 

\begin{figure}
\begin{center}
\resizebox{2.3in}{!}{\rotatebox{0}{\includegraphics{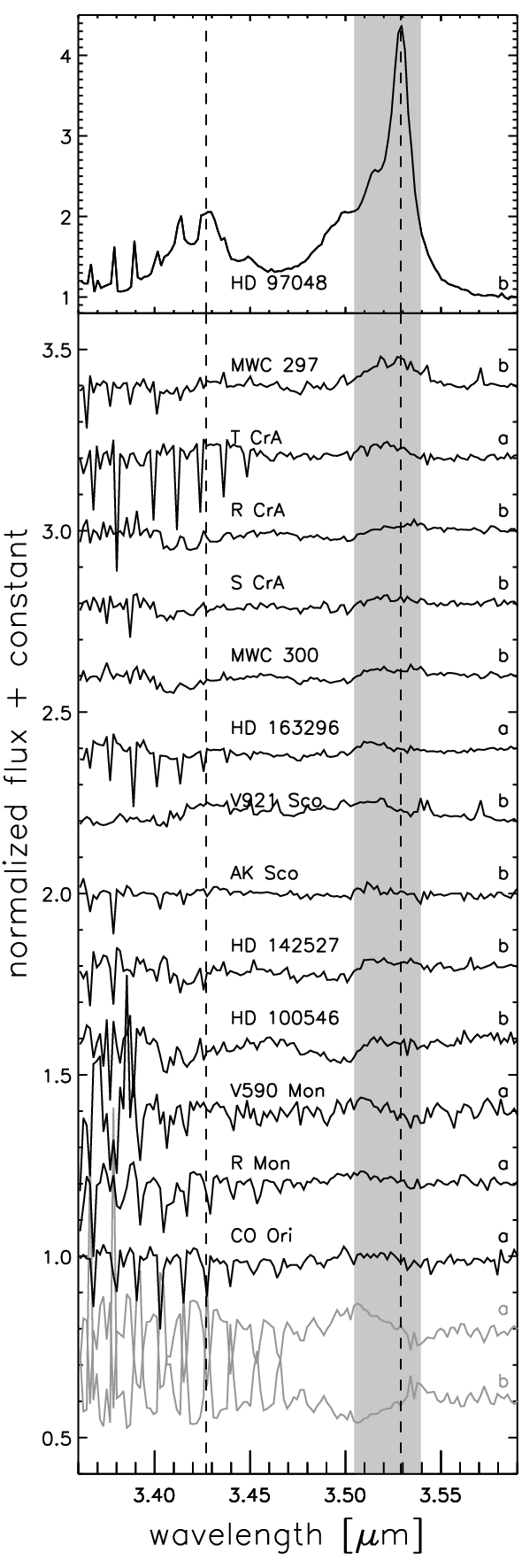}}} 
\caption{ The normalized spectra of the candidate nanodiamond
  emitters. The bottom two spectra are a standard-star spectrum, and
  the inverse of it. They represent the spectral shape of
  an undercorrected continuum $e^{-\tau_\lambda}$ (a) and an
  overcorrected continuum $e^{\tau_\lambda}$ (b). The peak position of
  the two curves are indicated by the edges of the grey box. The target
  spectra are flagged with a and b, based on the apparent shape of the
  continuum. The real detected nanodiamond features in MWC~297 and
  HD~97048 are plotted at the top of the figure. The latter is drawn
  in a separate box because of its impressive strength compared to the
  features or artefacts in the other spectra. The peak position of the
  3.53 and 3.43~$\mu$m features in this spectrum have been indicated by
  the dashed line. Most of the detected features can be brought back
  to data reduction artefacts. Only the features in V921~Sco,
  HD~163296 and T~CrA may have deviating shapes.}
\label{nanocans.ps}
\end{center}
\end{figure}

In our opinion, the main source of
data reduction artefacts is the correction procedure for atmospheric
absorption lines. Assume that $S_\lambda$ is the flux emitted by the
standard star (STD), and $\tau_\lambda$ the atmospheric optical depth
at a certain airmass and at the time of the STD
observation. Define the scientific target's flux as $T_\lambda$ and
the corresponding atmospheric optical depth as $\sigma_\lambda$. We
assume that the airmasses at the time of both observations are the
same. Note however that even in this case $\tau_\lambda$ and
$\sigma_\lambda$ are not necessarily the same, since the atmospheric
conditions (e.g. air density, humidity) may differ between the two
observations. The STD's flux can be approximated by a
photospheric model. If $F_\lambda$ is the measured STD flux, then
$e^{\tau_\lambda} = S_{\lambda,model} / F_\lambda$ can be
estimated. This atmospheric correction factor is then applied to the
target measurement, which leads to an estimate of the target's flux,
$\widetilde{T}_\lambda = (T_\lambda\ e^{-\sigma_{\lambda}})\
e^{\tau_{\lambda}} = T_\lambda\ e^{(\tau_\lambda - \sigma_\lambda)}
\approx T_\lambda$. Slight differences between $\tau_\lambda$ and
$\sigma_\lambda$ introduce artefacts in the reduced spectrum of the
science target.  As a first order approach, we assume that the main
difference in telluric extinction occurs due to a geometrical and/or
atmospheric density effect. Such an effect is linear in
$\tau_\lambda$. Hence $e^{(\tau_\lambda - \sigma_\lambda)}$ is
proportional to $e^{C\tau_\lambda}$, with $C$ a constant. The bottom
two spectra in Fig.~\ref{nanocans.ps} show the influence of {\em
over-} and {\em undercorrection} on a flat continuum. The artefacts
that are relevant in this discussion are the ``emission features'' at
3.505~$\mu$m in the case of undercorrection ($\propto
e^{-\tau_\lambda}$), and 3.54~$\mu$m in the case of overcorrection
($\propto e^{\tau_\lambda}$).

We will try to separate the real spectral
features from the artefacts in the 14 candidate nanodiamond emitters
listed in Table~\ref{presoffeat}. Two of them, HD~97048 and MWC~297,
are known to display the 3.53~$\mu$m feature and will serve as a
reference. The vast majority of the other 12 detected features is most
likely due to the data reduction procedure. In Fig.~\ref{nanocans.ps}, the
normalized spectra of the candidate nanodiamond emitters have been
displayed. We have indicated which spectra appear to be
over- and undercorrected. Most of
the features follow the spectral shape of $e^{\tau_\lambda}$ or
$e^{-\tau_\lambda}$. Only for V921~Sco, HD 163296 and T~CrA, the
detected feature's appearance deviates from the artefacts' shape. Even
in these cases however, it remains difficult to exclude an artefactual
origin. An indication may come from the upper limit to the 3.53~$\mu$m
line fluxes measured in the ISO spectrum. The upper limits for
HD 163296 and T~CrA are in agreement with the measured
ISAAC line fluxes ($<$2.5 and $<$1.4 vs. 2.3 and 0.65
$\times$~10$^{-15}$ W\,m$^{-2}$). For V921~Sco, the observed line flux
exceeds the ISO upper limit by a factor three ($<$3.7 vs. 10.3
$\times$~10$^{-15}$ W\,m$^{-2}$), which makes the NANO 3.53~$\mu$m
detection in this source rather suspicious. In the paper,
V921~Sco, HD 163296 and T~CrA are nevertheless treated as the
remaining three candidate nanodiamond emitters.

\acknowledgements{ The authors thank R.~van~Boekel for
  kindly providing us with the TIMMI2 spectrum of V921~Sco. Atomic
  data compiled in the DREAM data base \citep{biemont99} were
  extracted via VALD \citep[][and references therein]{kupka99}.}


\bibliographystyle{aa}
\bibliography{/STER/55/bram/REFERENCES/references.bib}

\begin{longtable}{rlrccccccl}
\caption{ The observation log of the sample stars
  The right ascension (RA), declination (Dec) and date of the ISAAC
  observations are indicated. The date, UT and
  integration time T of the observation are given. We have assigned a
  quality label (Q) to the ISAAC spectra: G--good quality; L--low flux
  levels ($<$0.5~Jy); J--large flux-level jump ($>$50\%) between the
  3.3 and 3.5~micron spectrum. The seeing at $\lambda$=0.5~$\mu$m, 
  estimated by the 
  VLT Astronomical Site Monitor (ASM), is given at the time of the 3.3 
  and 3.5~$\mu$m ISAAC observations respectively. The final column lists 
  the spectral calibrator stars. These standard stars have been observed 
  during the same night as the science target, unless when flagged 
  with $\star$.
\label{obslog}}   \\
\hline\hline
 \multicolumn{1}{r}{\#} &
 \multicolumn{1}{l}{Target} &
 \multicolumn{1}{c}{RA (2000)} &
 \multicolumn{1}{c}{Dec (2000)} &
 \multicolumn{1}{c}{Date} &
 \multicolumn{1}{c}{UT} &
 \multicolumn{1}{c}{T} &
 \multicolumn{1}{c}{Q} &
 \multicolumn{1}{c}{Seeing} &
 \multicolumn{1}{l}{Calibrator} \\
  &  & hh:mm:ss & dd:mm:ss & dd/mm/yy & hh:mm & m  &  & arcsec &   \\
\hline
\endfirsthead
\caption{ --- (Continued)}  \\
\hline\hline
 \multicolumn{1}{r}{\#} &
 \multicolumn{1}{l}{Target} &
 \multicolumn{1}{c}{RA (2000)} &
 \multicolumn{1}{c}{Dec (2000)} &
 \multicolumn{1}{c}{Date} &
 \multicolumn{1}{c}{UT} &
 \multicolumn{1}{c}{T} &
 \multicolumn{1}{c}{Q} &
 \multicolumn{1}{c}{Seeing} &
 \multicolumn{1}{l}{Calibrator} \\
  &  & hh:mm:ss & dd:mm:ss & dd/mm/yy & hh:mm & m  &  & arcsec &  \\
\hline
\endhead
\hline
\endfoot
 1 &\object{UX Ori}     & 05:04:29.99 & $-$03:47:14.3 & 22/10/2003 & 08:38 & 10 & G    &   1.19\,/\,1.16  & $\gamma$~Ori  \\
 2 &\object{HD 34282}   & 05:16:00.48 & $-$09:48:35.4 & 05/11/2003 & 06:15 & 14 & G    &   1.10\,/\,1.40  & $\gamma$~Ori$^\star$  \\
 3 &\object{V346 Ori}   & 05:24:42.80 & $+$01:43:48.3 & 05/11/2003 & 07:13 &  6 & G    &   1.53\,/\,1.49  & $\gamma$~Ori$^\star$  \\
 4 &\object{CO Ori}     & 05:27:38.34 & $+$11:25:38.9 & 28/10/2004 & 04:37 &  8 & G    &   0.44\,/\,0.63  & HD~212330  \\
 5 &\object{HD 35929}   & 05:27:42.79 & $-$08:19:38.4 & 03/10/2004 & 07:38 & 18 & G    &   1.08\,/\,0.85  & HD~38858  \\
 6 &\object{St\Ha 41}   & 05:29:11.44 & $-$06:08:05.4 & 03/10/2004 & 08:56 &  8 & G    &   1.08\,/\,1.08  & HD~38858  \\
 7 &\object{HK Ori}	& 05:31:28.01 & $+$12:09:10.7 & 09/02/2004 & 00:36 &  6 & J    &   1.17\,/\,0.84  & $\gamma$~Ori  \\
 8 &\object{HD 244604}  & 05:31:57.25 & $+$11:17:41.4 & 09/02/2004 & 01:39 &  6 & G    &   0.88\,/\,1.07  & $\gamma$~Ori$^\star$  \\
 9 &\object{RY Ori}     & 05:32:09.96 & $-$02:49:46.8 & 04/10/2004 & 06:34 & 18 & G    &   0.57\,/\,0.60  & HD~44594  \\
10 &\object{T Ori}      & 05:35:50.35 & $-$05:28:35.1 & 05/11/2003 & 07:39 &  6 & G    &   1.72\,/\,1.66  & $\gamma$~Ori$^\star$  \\
11 &\object{V380 Ori}   & 05:36:25.43 & $-$06:42:57.7 & 05/11/2003 & 08:00 &  6 & G    &   1.64\,/\,1.60  & $\gamma$~Ori$^\star$  \\
12 &\object{V586 Ori}   & 05:36:59.25 & $-$06:09:16.4 & 05/11/2003 & 08:27 & 18 & G    &   1.98\,/\,1.71  & $\gamma$~Ori$^\star$  \\
13 &\object{BF Ori}     & 05:37:13.26 & $-$06:35:00.6 & 08/10/2004 & 06:51 & 18 & G    &   1.46\,/\,1.64  & $\gamma$~Ori  \\
14 &\object{HD 37357}   & 05:37:47.08 & $-$06:42:30.3 & 08/10/2004 & 07:24 & 14 & G    &   1.56\,/\,1.89  & $\gamma$~Ori  \\
15 &\object{N3sk 81}    & 05:38:09.30 & $-$06:49:17.0 & 08/10/2004 & 09:06 & 14 & G    &   1.69\,/\,1.86  & $\gamma$~Ori  \\
16 &\object{HD 37411}   & 05:38:14.51 & $-$05:25:13.3 & 08/10/2004 & 08:38 & 10 & G    &   1.31\,/\,1.18  & $\gamma$~Ori  \\
17 &\object{Haro 13a}   & 05:38:18.20 & $-$07:02:26.6 & 28/10/2004 & 05:39 &  6 & G    &   0.42\,/\,0.44  & HD~38858  \\
18 &\object{V599 Ori}   & 05:38:58.60 & $-$07:16:46.0 & 09/10/2004 & 08:36 & 12 & G    &   0.61\,/\,0.76  & $\epsilon$~Cen$^\star$  \\
19 &\object{V350 Ori}   & 05:40:11.78 & $-$09:42:11.4 & 09/10/2004 & 05:05 & 18 & G    &   1.13\,/\,1.48  & $\epsilon$~Cen$^\star$  \\
20 &\object{HD 37806}   & 05:41:02.29 & $-$02:43:00.7 & 28/10/2004 & 06:46 &  6 & G    &   0.48\,/\,0.45  & HD~38858  \\
21 &\object{HD 250550}  & 06:01:59.99 & $+$16:30:56.7 & 12/01/2005 & 04:29 &  6 & G    &   1.78\,/\,1.75  & $\gamma$~Ori  \\
22 &\object{VY Mon}     & 06:31:06.90 & $+$10:26:05.3 & 12/01/2005 & 04:50 &  4 & G    &   2.08\,/\,2.00  & $\gamma$~Ori  \\
23 &\object{Lk\Ha 215}  & 06:32:41.80 & $+$10:09:33.6 & 21/01/2005 & 01:55 &  8 & G    &   0.76\,/\,0.70  & $\xi$~Ori  \\
   &                    & 	      &               & 26/02/2005 & 02:08 &  8 &      &   1.20\,/\,1.42  & $\rho$~Cen  \\
24 &\object{HD 259431}  & 06:33:05.19 & $+$10:19:20.0 & 21/01/2005 & 02:54 &  6 & G    &   0.88\,/\,0.79  & $\lambda$~Lep  \\
   &                    &             &               & 25/02/2005 & 03:42 &  6 &      &   0.67\,/\,0.57  & $\epsilon$~Cen  \\
25 &\object{R Mon}      & 06:39:09.89 & $+$08:44:10.0 & 10/01/2005 & 05:32 &  6 & G    &   0.96\,/\,0.72  & $\gamma$~Ori  \\
26 &\object{V590 Mon}   & 06:40:44.57 & $+$09:48:02.1 & 10/01/2005 & 03:46 & 28 & G    &   0.73\,/\,1.94  & $\gamma$~Ori  \\
27 &\object{GU CMa}     & 07:01:49.51 & $-$11:18:03.3 & 01/03/2004 & 03:48 &  6 & J    &   0.66\,/\,0.84  & $\epsilon$~Cen  \\
28 &\object{HD 53367}   & 07:04:25.53 & $-$10:27:15.7 & 12/02/2005 & 05:45 &  6 & G    &   1.00\,/\,0.76  & $\beta$~Cir  \\
   &                    &             &               & 19/11/2003 & 08:15 &  6 &      &   0.50\,/\,0.51  & $\gamma$~Ori$^\star$  \\
29 &\object{NX Pup}     & 07:19:28.26 & $-$44:35:11.3 & 09/02/2004 & 05:04 &  6 & G    &   0.81\,/\,0.85  & $\gamma$~Ori  \\
30 &\object{HD 76534}   & 08:55:08.71 & $-$43:27:59.9 & 11/01/2005 & 04:54 & 18 & G    &   2.03\,/\,1.85  & $\nu$~Pup  \\
31 &\object{HD 85567}   & 09:50:28.54 & $-$60:58:03.0 & 20/01/2005 & 09:06 &  6 & G    &   0.64\,/\,0.59  & $\pi$~Sco  \\
32 &\object{HD 95881}   & 11:01:57.62 & $-$71:30:48.4 & 23/01/2005 & 05:34 &  6 & G    &   0.81\,/\,0.76  & $\xi^2$~Cen  \\
33 &\object{HD 97048}   & 11:08:03.32 & $-$77:39:17.5 & 18/07/2004 & 01:01 &  6 & G    &   0.55\,/\,0.63  & $\epsilon$~Cen$^\star$  \\
34 &\object{HD 100546}  & 11:33:25.44 & $-$70:11:41.2 & 27/07/2004 & 23:49 &  6 & G    &   1.24\,/\,1.41  & $\kappa$~Eri  \\
35 &\object{HD 101412}  & 11:39:44.46 & $-$60:10:27.7 & 28/07/2004 & 00:20 &  8 & G    &   1.33\,/\,1.29  & $\kappa$~Eri  \\
36 &\object{T Cha}	& 11:57:13.53 & $-$79:21:31.5 & 18/07/2004 & 01:26 & 10 & J    &   0.88\,/\,0.71  & $\epsilon$~Cen$^\star$  \\
37 &\object{LSS 3027 B} & 13:19:03.98 & $-$62:34:10.1 & 10/05/2004 & 05:39 &  8 & L    &   0.78\,/\,0.90  & $\epsilon$~Cen  \\
38 &\object{HD 130437}  & 14:50:50.26 & $-$60:17:10.3 & 10/05/2004 & 06:03 &  8 & G    &   0.93\,/\,0.84  & $\epsilon$~Cen  \\
39 &\object{SS73 44}    & 15:03:23.80 & $-$63:22:59.0 & 05/09/2003 & 01:08 &  8 & G    &   0.83\,/\,0.95  & $\kappa$~Cet$^\star$  \\
40 &\object{HD 132947}  & 15:04:56.05 & $-$63:07:52.6 & 05/09/2003 & 01:35 &  8 &L$+$J &   0.88\,/\,0.93  & $\kappa$~Cet$^\star$ \\
41 &\object{HD 141569}  & 15:49:57.75 & $-$03:55:16.4 & 22/08/2004 & 23:15 &  8 & J    &   0.72\,/\,0.73  & $\sigma$~Sco  \\
42 &\object{HD 142666}  & 15:56:40.02 & $-$22:01:40.0 & 23/08/2004 & 00:53 &  6 & J    &   0.96\,/\,0.96  & $\sigma$~Sco  \\
43 &\object{HD 142527}  & 15:56:41.89 & $-$42:19:23.3 & 05/09/2003 & 02:05 &  6 & G    &   1.32\,/\,0.91  & $\kappa$~Cet$^\star$  \\
44 &\object{HR 5999}    & 16:08:34.29 & $-$39:06:18.3 & 22/07/2004 & 03:46 &  6 & J    &   1.61\,/\,1.82  & HD~172910  \\
45 &\object{HD 150193}  & 16:40:17.92 & $-$23:53:45.2 & 10/05/2004 & 06:50 &  6 & G    &   0.83\,/\,0.86  & $\kappa$~Aql  \\
46& \object{AK Sco}	& 16:54:44.85 & $-$36:53:18.6 & 10/05/2004 & 07:15 &  8 & G    &   0.85\,/\,0.89  & $\kappa$~Aql  \\
47& \object{V921 Sco}   & 16:59:06.90 & $-$42:42:08.0 & 22/07/2004 & 03:01 &  6 & G    &   1.76\,/\,1.82  & HD~172910  \\
48& \object{KK Oph}	& 17:10:08.07 & $-$27:15:18.2 & 22/07/2004 & 03:26 &  8 & J    &   1.39\,/\,1.36  & HD~172910  \\
49& \object{HD 163296}  & 17:56:21.29 & $-$21:57:21.9 & 19/09/2003 & 02:07 &  6 & G    &   1.54\,/\,1.74  & $\kappa$~Cet  \\
50& \object{MWC 297}    & 18:27:39.60 & $-$03:49:52.0 & 19/09/2003 & 02:27 &  4 & G    &   1.67\,/\,1.64  & $\kappa$~Cet  \\
51& \object{VV Ser}	& 18:28:47.87 & $+$00:08:39.6 & 19/09/2003 & 02:46 &  6 & G    &   1.42\,/\,1.32  & $\kappa$~Cet  \\
52& \object{MWC 300}    & 18:29:25.69 & $-$06:04:37.3 & 19/09/2003 & 03:02 &  6 & G    &   1.42\,/\,1.45  & $\kappa$~Cet  \\
53& \object{AS 310}	& 18:33:21.17 & $-$04:58:06.7 & 19/09/2003 & 03:26 & 14 & L    &   1.57\,/\,1.28  & $\kappa$~Cet  \\
54& \object{S CrA}	& 19:01:08.60 & $-$36:57:20.0 & 19/09/2003 & 03:48 &  8 & G    &   1.49\,/\,1.39  & $\kappa$~Cet  \\
55& \object{TY CrA}	& 19:01:40.83 & $-$36:52:33.9 & 19/09/2003 & 04:08 &  8 & G    &   1.14\,/\,1.04  & $\kappa$~Cet  \\
56& \object{R CrA}	& 19:01:53.65 & $-$36:57:07.6 & 19/09/2003 & 04:27 &  6 & G    &   0.83\,/\,0.91  & $\kappa$~Cet  \\
57& \object{T CrA}	& 19:01:58.80 & $-$36:57:49.0 & 26/09/2003 & 01:38 &  8 & G    &   0.99\,/\,1.24  & BY~Cap  \\
58& \object{HD 179218}  & 19:11:11.25 & $+$15:47:15.6 & 01/10/2003 & 02:01 &  6 & G    &   0.63\,/\,0.66  & HD~9562  \\
59& \object{HBC 687}    & 19:29:00.80 & $+$09:38:43.0 & 26/09/2003 & 02:08 & 20 & L    &   1.11\,/\,1.12  & HD~9562  \\
60& \object{V1295 Aql}  & 20:03:02.52 & $+$05:44:16.7 & 19/09/2003 & 04:46 &  6 & G    &   0.78\,/\,0.90  & $\kappa$~Cet  \\
\end{longtable}		

\begin{longtable}{rlr|cccc|cccc}
\caption{ The detected features in the 3~micron region of the sample
  stars. The sample stars have been classified in
  Meeus groups~I and II \citep{meeus01}. Targets with $A_V$ exceeding 4
  are classified as embedded sources (E). Sources with unknown or
  uncertain classification are flagged with ``?''. The measured line fluxes 
  and peak-over-continuum ratios of four spectral features are given: the 
  Pfund~$\delta$ emission line at 3.296~$\mu$m, the PAH 3.3~$\mu$m
  feature and the nanodiamond features at 3.43 and
  3.53~$\mu$m. Tentative detections (line flux between 1$-$3$\sigma$) are
  indicated with~$a$. The flag~$b$ indicates 
  detected features of unclear origin, either linked to data reduction
  artefacts, nanodiamonds or an unknown carrier. 
\label{presoffeat}}   \\
\hline\hline
 \multicolumn{1}{r}{\#} &
 \multicolumn{1}{l}{Target} &
 \multicolumn{1}{r|}{Group} &
 \multicolumn{1}{|c}{Pf~$\delta$} &
 \multicolumn{1}{c}{PAH} &
 \multicolumn{1}{c}{NANO} &
 \multicolumn{1}{c|}{NANO} & 
 \multicolumn{1}{|c}{Pf~$\delta$} &
 \multicolumn{1}{c}{PAH} &
 \multicolumn{1}{c}{NANO} &
 \multicolumn{1}{c}{NANO} \\ 
  & &  & 3.296~$\mu$m & 3.3~$\mu$m & 3.43~$\mu$m & 3.53~$\mu$m & 3.296~$\mu$m & 3.3~$\mu$m & 3.43~$\mu$m & 3.53~$\mu$m \\
 \multicolumn{3}{c|}{ } &
 \multicolumn{4}{|c|}{line flux$/10^{-16}$~\Wm} &
 \multicolumn{4}{|c}{peak-over-continuum ratio} \\
\hline
\endfirsthead
\caption{ --- (Continued)}  \\
\hline\hline
 \multicolumn{1}{r}{\#} &
 \multicolumn{1}{l}{Target} &
 \multicolumn{1}{r|}{Group} &
 \multicolumn{1}{|c}{Pf~$\delta$} &
 \multicolumn{1}{c}{PAH} &
 \multicolumn{1}{c}{NANO} &
 \multicolumn{1}{c|}{NANO} & 
 \multicolumn{1}{|c}{Pf~$\delta$} &
 \multicolumn{1}{c}{PAH} &
 \multicolumn{1}{c}{NANO} &
 \multicolumn{1}{c}{NANO} \\ 
  & &  & 3.296~$\mu$m & 3.3~$\mu$m & 3.43~$\mu$m & 3.53~$\mu$m & 3.296~$\mu$m & 3.3~$\mu$m & 3.43~$\mu$m & 3.53~$\mu$m \\
 \multicolumn{3}{c|}{ } &
 \multicolumn{4}{|c|}{line flux$/10^{-16}$~\Wm} &
 \multicolumn{4}{|c}{peak-over-continuum ratio} \\
\hline
\endhead
\hline
\endfoot
 1 &\object{UX Ori}     &   II &   0.7 $\pm$   0.2 	&         $<$    14      &   $<$   16    &        $<$     2.9 	             &      1.18 $\pm$     0.05	& $<$  1.05 		 & $<$ 1.05            & $<$ 1.01     	        \\
 2 &\object{HD 34282}   &    I &        $<$    1.1   	&     24  $\pm$  7       &   $<$   17    &        $<$     4.7 	  	     & $<$  1.06 		&      1.22 $\pm$     0.07	 & $<$ 1.06	     & $<$ 1.02	        \\
 3 &\object{V346 Ori}   &    I &        $<$    1.4   	&     17  $\pm$  8$^{\,a}$   &   $<$   23    &        $<$     8.4 	     & $<$  1.10 		&      1.2  $\pm$     0.1	 & $<$ 1.10	     & $<$ 1.05	        \\
 4 &\object{CO Ori}     &   II &        $<$    0.4   	&         $<$    6.9     &   $<$    8.2  &   1.2  $\pm$   0.7$^{\,a,b}$      & $<$  1.05 		& $<$  1.05 		 & $<$ 1.15          &    1.02 $\pm$ 0.02     	\\
 5 &\object{HD 35929}   &   II &   0.4 $\pm$   0.4$^{\,a}$ 	&         $<$    12      &   $<$   14    &        $<$     2.7 	     &      1.04 $\pm$     0.04	& $<$  1.04 		 & $<$ 1.04	     & $<$ 1.01	             	\\
 6 &\object{St\Ha 41}   &  II? &        $<$   0.32   	&         $<$    5.8     &   $<$   10    &        $<$     4.2 	  	     & $<$  1.05 		& $<$  1.05 		 & $<$ 1.15          & $<$ 1.05	      	     	\\
 7 &\object{HK Ori}	&   II &        $<$    1.1   	&         $<$    31      &   $<$   35    &        $<$     9.0 	  	     & $<$  1.08 		& $<$  1.08 		 & $<$ 1.08	     & $<$ 1.03	             	\\
 8 &\object{HD 244604}  &   II &        $<$    1.5   	&         $<$    27      &   $<$   31    &        $<$     8.0 	  	     & $<$  1.10 		& $<$  1.10 		 & $<$ 1.10	     & $<$ 1.04	             	\\
 9 &\object{RY Ori}     &   II &       $<$    0.25   	&         $<$    4.8     &   $<$    5.7  &        $<$     1.1 	  	     & $<$  1.06 		& $<$  1.06 		 & $<$ 1.06	     & $<$ 1.02	             	\\
10 &\object{T Ori}      &  II? &       $<$     3.7   	&         $<$    62      &   $<$   69    &        $<$    12   	  	     & $<$  1.06 		& $<$  1.06 		 & $<$ 1.06	     & $<$ 1.01	             	\\
11 &\object{V380 Ori}   &  II? &   5   $\pm$       1 	&         $<$    86      &   $<$  110    &        $<$    22   	  	     &      1.27 $\pm$     0.08	& $<$  1.07 		 & $<$ 1.08	     & $<$ 1.02	             	\\
12 &\object{V586 Ori}   &   II &  0.25 $\pm$    0.09$^{\,a}$&         $<$    8.7     &   $<$   10    &        $<$     3.0 	     &      1.13 $\pm$     0.05	& $<$  1.05 		 & $<$ 1.05	     & $<$ 1.02	             	\\
13 &\object{BF Ori}     &   II &   0.6 $\pm$     0.2 	&         $<$    7.6     &   $<$    8.8  &        $<$     3.5 	  	     &      1.15 $\pm$     0.05	& $<$  1.05 		 & $<$ 1.05	     & $<$ 1.03	             	\\
14 &\object{HD 37357}   &   II &   2.1 $\pm$     0.4 	&         $<$    6.2     &   $<$    7.1  &        $<$     3.9 	  	     &      1.15 $\pm$     0.03	& $<$  1.03 		 & $<$ 1.03	     & $<$ 1.02	             	\\
15 &\object{N3sk 81}    &    E &       $<$    0.58   	&         $<$    9.7     &   $<$   11    &        $<$     3.1 	  	     & $<$  1.04         	& $<$  1.04 		 & $<$ 1.04	     & $<$ 1.02	             	\\
16 &\object{HD 37411}   &   II &       $<$    0.65   	&         $<$    10      &   $<$   12    &        $<$     4.4 	  	     & $<$  1.04 		& $<$  1.04 		 & $<$ 1.04	     & $<$ 1.02	             	\\
17 &\object{Haro 13a}   &    I &       $<$     1.9   	&         $<$    34      &   $<$   44    &        $<$     7.3 	  	     & $<$  1.03 		& $<$  1.03 		 & $<$ 1.03	     & $<$ 1.01	             	\\
18 &\object{V599 Ori}   &    E &       $<$    0.24   	&      5  $\pm$  1       &   $<$    4.2  &        $<$     0.95	  	     & $<$  1.04 		&      1.15 $\pm$     0.04     	 & $<$ 1.04	     & $<$ 1.01	        \\
19 &\object{V350 Ori}   &   II &       $<$   0.063   	&         $<$    1.1     &   $<$    1.2  &        $<$     0.37	  	     & $<$  1.06 	        & $<$  1.06 		 & $<$ 1.06	     & $<$ 1.02	             	\\
20 &\object{HD 37806}   &   II &   1.3 $\pm$     0.3 	&         $<$    17      &   $<$   20    &        $<$     2.3 	  	     &      1.14 $\pm$     0.03	& $<$  1.03 		 & $<$ 1.03	     & $<$ 1.01	             	\\
21 &\object{HD 250550}  &    I &       $<$     3.3   	&         $<$    59      &   $<$   69    &        $<$     8.3 	  	     & $<$  1.16 		& $<$  1.16 		 & $<$ 1.16	     & $<$ 1.03	             	\\
22 &\object{VY Mon}     &    E &       $<$      31   	&         $<$    530     &   $<$  670    &        $<$    38   	  	     & $<$  1.17 		& $<$  1.17 		 & $<$ 1.17	     & $<$ 1.01	             	\\
23 &\object{Lk\Ha 215}  &    I &       $<$    0.40   	&         $<$    7.3     &   $<$    8.8  &        $<$     1.7 	  	     & $<$  1.07 		& $<$  1.07 		 & $<$ 1.07	     & $<$ 1.02	             	\\
24 &\object{HD 259431}  &    I &   0.6 $\pm$     0.2$^{\,a}$&         $<$    32      &   $<$   37    &        $<$     4.5            &      1.16 $\pm$     0.06	& $<$  1.06 		 & $<$ 1.06	     & $<$ 1.01	             	\\
25 &\object{R Mon}      &    I &       $<$     13    	&         $<$    260     &   $<$  350    &   28   $\pm$  10$^{\,a,b}$        & $<$  1.08 		& $<$  1.08 		 & $<$ 1.08	     &    1.03 $\pm$ 0.01      	\\
26 &\object{V590 Mon}   &    ? &       $<$     1.9   	&      7  $\pm$    6$^{\,a}$ &   $<$   34    &   2.4  $\pm$   0.8$^{\,a,b}$  & $<$  1.16 		&      1.2  $\pm$     0.2      	 & $<$ 1.16	     &    1.05 $\pm$ 0.02 \\
27 &\object{GU CMa}     &   II &   4.9 $\pm$     0.8 	&         $<$    43      &   $<$   48    &        $<$    12   	  	     &      1.46 $\pm$     0.09	& $<$  1.08 		 & $<$ 1.08	     & $<$ 1.03	             	\\
28 &\object{HD 53367}   &    ? &       $<$     1.3   	&         $<$    20      &   $<$   24    &        $<$     5.1 	  	     & $<$  1.11 		& $<$  1.11 		 & $<$ 1.11	     & $<$ 1.03	             	\\
29 &\object{NX Pup}     &   II &       $<$     3.4   	&         $<$    67      &   $<$   75    &        $<$    15   	  	     & $<$  1.04 		& $<$  1.04 		 & $<$ 1.04	     & $<$ 1.01	             	\\
30 &\object{HD 76534}   &    ? &       $<$    0.27   	&         $<$    4.5     &   $<$    5.2  &        $<$     0.59	  	     & $<$  1.19 		& $<$  1.19 		 & $<$ 1.19	     & $<$ 1.03	             	\\
31 &\object{HD 85567}   &   II &       $<$     1.2   	&         $<$    19      &   $<$   21    &        $<$     3.0 	  	     & $<$  1.06 		& $<$  1.06 		 & $<$ 1.06	     & $<$ 1.01	             	\\
32 &\object{HD 95881}   &   II &       $<$     3.1   	&         $<$    49      &   $<$   56    &        $<$     9.9 	  	     & $<$  1.10 		& $<$  1.10 		 & $<$ 1.10	     & $<$ 1.02	             	\\
33 &\object{HD 97048}   &    I &       $<$     1.5   	&     65  $\pm$  20      &200 $\pm$ 20   & 486    $\pm$   5   	  	     & $<$  1.10 		&      1.4  $\pm$     0.1      	 &    2.0 $\pm$ 0.1    &    4.3 $\pm$ 0.2 \\
34 &\object{HD 100546}  &    I &   9   $\pm$       2 	&    240  $\pm$  40      &   $<$   94    &  13    $\pm$   10$^{\,a,b}$       &      1.33 $\pm$     0.06	&      1.34 $\pm$     0.06     	 & $<$ 1.06	     &    1.05 $\pm$ 0.04 \\
35 &\object{HD 101412}  &   II &       $<$     1.3   	&     13  $\pm$  10$^{\,a}$  &   $<$   25    &        $<$     7.5 	     & $<$  1.06 		&      1.08 $\pm$     0.06     	 & $<$ 1.06	     & $<$ 1.03	        \\
36 &\object{T Cha}	&   II &       $<$    0.85   	&         $<$    16      &   $<$   17    &        $<$     4.0 	  	     & $<$  1.07 		& $<$  1.07 		 & $<$ 1.07	     & $<$ 1.02	             	\\
37 &\object{LSS 3027 B} &    ? &       $<$    0.46   	&         $<$    8.7     &   $<$    8.8  &        $<$     3.6 	  	     & $<$  1.73 		& $<$  1.73 		 & $<$ 1.73	     & $<$ 1.39	             	\\
38 &\object{HD 130437}  &    ? &   1.8 $\pm$     0.2 	&         $<$    6.5     &   $<$    7.1  &        $<$     5.6 	  	     &      1.64 $\pm$     0.06	& $<$  1.06 		 & $<$ 1.06	     & $<$ 1.06	             	\\
39 &\object{SS73 44}    &    I &       $<$    0.89   	&         $<$    16      &   $<$   17    &        $<$     8.8 	  	     & $<$  1.05 		& $<$  1.05 		 & $<$ 1.05	     & $<$ 1.04	             	\\
40 &\object{HD 132947}  &   II &       $<$    0.68   	&         $<$    8.6     &   $<$    9.1  &        $<$     8.1 	  	     & $<$  1.14 		& $<$  1.14 		 & $<$ 1.14	     & $<$ 1.18	             	\\
41 &\object{HD 141569}  &   II &       $<$    0.87   	&         $<$    37      &   $<$   38    &        $<$     5.9 	  	     & $<$  1.20 		& $<$  1.20 		 & $<$ 1.20	     & $<$ 1.05	             	\\
42 &\object{HD 142666}  &   II &       $<$    0.72   	&         $<$    20      &   $<$   21    &        $<$     5.8 	  	     & $<$  1.04 		& $<$  1.04 		 & $<$ 1.04	     & $<$ 1.02	             	\\
43 &\object{HD 142527}  &    I &    32 $\pm$       5 	&         $<$    110     &   $<$  120    &  44    $\pm$   9$^{\,b}$   	     &      1.27 $\pm$     0.04	& $<$  1.04 		 & $<$ 1.04	     &     1.05 $\pm$ 0.01      	\\
44 &\object{HR 5999}    &   II &   5   $\pm$       3$^{\,a}$&         $<$    110     &   $<$  150    &        $<$    55              &      1.08 $\pm$     0.04	& $<$  1.04 		 & $<$ 1.04	     & $<$ 1.02	             	\\
45 &\object{HD 150193}  &   II &   3.4 $\pm$     0.3 	&         $<$    13      &   $<$   15    &        $<$     6.3 	  	     &      1.15 $\pm$     0.01	& $<$  1.01 		 & $<$ 1.01	     & $<$ 1.01	             	\\
46 &\object{AK Sco}	&   II &       $<$    0.79   	&         $<$    13      &   $<$   15    &   2.5  $\pm$   0.9$^{\,a,b}$      & $<$  1.03 		& $<$  1.03 		 & $<$ 1.03	     &    1.03 $\pm$ 0.01      	\\
47 &\object{V921 Sco}   &    E &    86 $\pm$       5 	&         $<$    160     &   $<$  190    & 103    $\pm$  30$^{\,b}$   	     &      1.42 $\pm$     0.03	& $<$  1.03 		 & $<$ 1.08          &    1.04 $\pm$ 0.01     	\\
48 &\object{KK Oph}	&   II &       $<$     2.8   	&         $<$    47      &   $<$   66    &        $<$    19   	  	     & $<$  1.05 		& $<$  1.05 		 & $<$ 1.05	     & $<$1.02	             	\\
49 &\object{HD 163296}  &   II &   18  $\pm$       3 	&         $<$    120     &   $<$  130    &  23    $\pm$   7$^{\,b}$   	     &      1.16 $\pm$     0.03	& $<$  1.03 		 & $<$ 1.03	     &    1.03 $\pm$ 0.01      	\\
50 &\object{MWC 297}    &    E &   240 $\pm$      20 	&         $<$    910     &   $<$ 1100    & 535    $\pm$ 100   	  	     &      1.63 $\pm$     0.05	& $<$  1.04 		 & $<$ 1.04	     &    1.06 $\pm$ 0.01      	\\
51 &\object{VV Ser}	&   II &   2.1 $\pm$     0.4 	&         $<$    23      &   $<$   25    &        $<$     9.0 	  	     &      1.13 $\pm$     0.03	& $<$  1.03 		 & $<$ 1.03	     & $<$1.01	             	\\
52 &\object{MWC 300}    &    E &   9   $\pm$       1 	&         $<$    82      &   $<$   97    &  50    $\pm$   9$^{\,b}$   	     &      1.13 $\pm$     0.02	& $<$  1.02 		 & $<$ 1.02	     &    1.03 $\pm$ 0.01      	\\
53 &\object{AS 310}	&    I &       $<$    0.83   	&     11  $\pm$  3       &   $<$    9.5  &        $<$     4.0 	  	     & $<$  1.34 		&      2.1  $\pm$     0.5      	 & $<$ 1.34	     & $<$ 1.22	        \\
54 &\object{S CrA}	&    ? &   3.9 $\pm$     0.4 	&         $<$    32      &   $<$   35    &  10    $\pm$   3$^{\,b}$   	     &      1.20 $\pm$     0.03	& $<$  1.03 		 & $<$ 1.03	     &    1.03 $\pm$ 0.01      	\\
55 &\object{TY CrA}	&    ? &       $<$    0.75   	&     40  $\pm$  4       &   $<$   12    &        $<$     4.7 	  	     & $<$  1.03 		&      1.28 $\pm$     0.03     	 & $<$ 1.03	     & $<$ 1.02	        \\
56 &\object{R CrA}	&    I &       $<$      26   	&         $<$    470     &   $<$  550    & 290    $\pm$  50$^{\,b}$   	     & $<$  1.02 		& $<$  1.02 		 & $<$ 1.02	     &    1.04 $\pm$ 0.01      	\\
57 &\object{T CrA}	&  II? &       $<$     2.5   	&         $<$    44      &   $<$   96    &   7    $\pm$   1$^{\,b}$   	     & $<$  1.06 		& $<$  1.06 		 & $<$ 1.19          &    1.04 $\pm$ 0.01     	\\
58 &\object{HD 179218}  &    I &       $<$     3.1   	&    160  $\pm$  20      &   $<$   47    &        $<$   6   		     & $<$  1.03 		&      1.28 $\pm$     0.03     	 & $<$ 1.03	     & $<$ 1.04         \\
59 &\object{HBC 687}    &   I? &       $<$    0.14   	&    1.7$\pm$ 0.7$^{\,a}$    &   $<$    2.8  &        $<$     1.0 	     & $<$  1.07 		&      1.14 $\pm$     0.07     	 & $<$ 1.07	     & $<$ 1.03	        \\
60 &\object{V1295 Aql}  &   II &   4.8 $\pm$     0.6    &         $<$    40      &   $<$   44    &        $<$    16                  &      1.18 $\pm$     0.02	& $<$  1.02 		 & $<$ 1.02	     & $<$ 1.01                  \\
\end{longtable}

\end{document}